\documentclass[iop]{emulateapj}
\usepackage{mathptmx}

\begin{document}

\title{Photometric Reverberation Mapping of the Broad Emission Line Region in Quasars}
\author{Doron Chelouche\altaffilmark{1} and Eliran Daniel\altaffilmark{2}}
\altaffiltext{1} {Department of Physics, Faculty of Natural Sciences, University of Haifa, Haifa 31905, Israel; doron@sci.haifa.ac.il}
\altaffiltext{2} {Department of Physics, Technion, Haifa 32000, Israel; elirandviv@gmail.com }
\shortauthors{Chelouche \& Daniel}
\shorttitle{Photometric Reverberation Mapping of Quasar BLR}

\begin{abstract}

A method is proposed for measuring the size of the broad emission line region (BLR) in quasars using broadband photometric data. A feasibility study, based on numerical simulations, points to the advantages and pitfalls associated with this approach. The method is applied to a subset of the Palomar-Green quasar sample for which independent BLR size measurements are available. An agreement is found between the results of the photometric method and the spectroscopic reverberation mapping technique. Implications for the measurement of BLR sizes and black hole masses for numerous quasars in the era of large surveys are discussed.

\end{abstract}


\section{Introduction}

Black holes (BH) are physical entities that provide insight into the foundations of (quantum-) gravity theories. Such objects are characterized by three basic properties: mass, angular momentum, and charge. Among these properties the most fundamental is the mass, which characterizes all BHs. Angular momentum is only relevant for rotating (Kerr) BHs while charge is less relevant for astrophysical BHs. 

Currently, the best evidence for the existence of BH is astrophysical. In particular, a broad range of BH masses is implied by astronomical observations: from solar mass BHs as the end products of massive stars to supermassive BHs (SMBH) at the centers of galaxies with masses that may easily exceed a million solar masses  \citep{sal64,lyn69}.

Perhaps the first evidence for the presence of SMBH at the centers of galaxies is associated with quasars and other types of active galactic nuclei (to which we shall henceforth refer to, generally, as quasars). In those objects, gas from the medium surrounding the SMBH, that is able to lose angular momentum and spiral inwards, eventually accretes onto it. As the gas accretes,  its gravitational energy is effectively converted to radiation with most of the emitted flux originating from the immediate vicinity of the SMBH, where the gas reaches the highest temperatures \citep{ss73}.

The mass of SMBHs in small samples of low redshift non-active galaxies has been measured using, for example, maser- and stellar-kinematics \citep{miy95,gen97,geb00a}. For most low redshift galaxies, SMBH mass estimation is, however, indirect and involves scaling relations that were derived for small samples of objects for which direct mass measurements were possible \citep{mag98}. 

As one attempts to estimate the mass of black holes at higher redshifts, various complications arise: it is not clear that locally determined scaling laws apply also to higher redshift objects since galaxies and their environments evolve with cosmic time. Furthermore, high redshift objects discovered by flux-limited samples are often much brighter than their low-$z$ counterparts, and it is yet to be established whether extrapolated relations are at all relevant in that regime \citep{net03}. For these reasons, direct SMBH mass determination of intermediate- and high-redshift objects is the most reliable way to further our understanding of SMBH-galaxy formation and co-evolution, and to reach a reliable census of black holes in the universe. 

Direct mass determination of SMBHs at high redshifts is currently limited to quasars, which are among the brightest objects in the universe and are powered by the SMBH itself. While quasars certainly trace the galaxy population, they also present a particular epoch in a galaxy lifetime, in which the black hole is accreting gas, and is at a stage of rapid growth. Therefore, better understanding of the SMBH demography in low- and  high-$z$ quasars has far-reaching implications for galaxy/black-hole formation and co-evolution \citep{fer01,sh03,os04,hop06,kol06,sh06,gh06,woo08,tra10}. 

A reliable method to directly measure the mass of SMBHs in quasars involves the reverberation mapping of the broad emission line region (BLR) in those objects \citep{bah72,pet93}. This region is known to be photo-ionized by the central continuum source \citep{bal78}, and its emission, in the form of broad permitted emission lines, is seen to react to continuum variations. Studies have shown that the BLR lies at distances, $R_{\rm BLR}$, that are much larger than the SMBH Schwarzschild radius, and also larger than the size of the continuum emitting region, i.e., the accretion disk. That being said, the BLR lies at small enough distances so that its kinematics is primarily set by the SMBH, which has a prominent contribution to the gravitational potential on those scales. That this is the case may be  deduced from the velocity dispersion of the emitted gas, as inferred from the width of the broad emission lines, which is typically of order a few\,$\times 10^3\,{\rm km~s^{-1}}$, as well as from the fact that, in luminous quasars, line variations lag behind relatively rapid continuum variations with typical time delays, $t_{\rm lag}\sim R_{\rm BLR}/c\simeq 200$\,days, where $c$ is the speed of light \citep[and references therein]{kas00}. 

Technically, reverberation mapping involves multi-epoch spectroscopic observations so that continuum and emission line variations may be individually traced \citep{di93,net97,kas00}. Continuum and emission line light curves can then be used to infer the size of the broad emission line region using a cross-correlation analysis \citep[and \S2.2]{bm82}. By measuring $R_{\rm BLR}$ and the velocity dispersion of the gas (via spectral modeling of the emission line profile and by defining a measure of the velocity dispersion of the gas), the SMBH mass may be deduced, up to a geometrical factor of order unity \citep{pw99,kro01}. Mass determinations using this technique have been shown to be consistent with the results of other independent methods \citep{geb00,fer01,pet10}, and are thought to be accurate to within 0.5\,dex \citep{kro01}. 

Nowadays, reverberation mapping has a central role in determining SMBH masses in active galaxies and for shedding light on the physics of the central engine in these objects\footnote{It is worth noting that, in addition to SMBH mass measurement, the structure and geometry of the BLR may be probed by means of velocity-resolved reverberation mapping of the broad emission lines \citep{kor95,wan95,kol03,ben10}. Such investigations cannot be carried out using broadband photometric reverberation techniques, which do not resolve the emission lines, and will not be discussed here any further.}. Nevertheless, despite several decades of spectroscopic monitoring of quasars, reliable BLR sizes and corresponding SMBH mass measurements have been determined for only a few dozens of low-$z$ objects \citep{gas86,kor91,di93,kas00,pet04,kas05,pet05,ben09,pet10,bar11}. Using the results of reverberation mapping, various scaling relations were found and quantified including, for example, the $R_{\rm BLR}$ (or SMBH mass)  vs. quasar luminosity relation \citep{kor91,kas00,pet04,kas07,ben09}, the SMBH mass vs. the quasar's optical spectral properties relation \citep{gh05}, and the black hole-host galaxy bulge mass relation in quasars \citep{wan99}. 

Despite being derived from small samples of nearby objects, the aforementioned relations are often extrapolated  to reflect on the quasar population as a whole \citep{ves02,ves09}, sometimes with little justification \citep{net03,kas07,mcg08,z08}. Potential biases arising from uncertain extrapolations to high redshifts, or to different sub-classes of quasars, may crucially affect our understanding of black hole growth and structure formation over cosmic time \citep{kh00,net03,col04,pel07,tra10}. Clearly, an efficient way to implement the reverberation mapping technique is highly desirable as it would alleviate many of the uncertainties currently plaguing the field.

Motivated by upcoming photometric surveys that will cover a broad range of wavelengths and will regularly monitor a fair fraction of the sky with good photometric accuracy (e.g., the {\it Large Synoptic Survey Telescope}, LSST), we aim to provide proof-of-concept that photometric reverberation mapping of the BLR in quasars is feasible. In a nut shell, instead of spectrally separating line and continuum light curves from multi-epoch spectroscopic observations, we take  advantage of the different variability properties of continuum and line processes and separate them at the light curve level. In this work we have no intention to address the full scope of issues associated with time-series analysis but rather hope to trigger further observational and theoretical work in broadband reverberation mapping of quasar BLRs.  

This paper is organized as follows: section 2 presents the photometric approach to reverberation mapping using analytic means and by resorting to simulations of quasar light curves. The reliability of the photometric method for line-to-continuum time lag measurement is investigated by means of simulations in \S3. In \S4 we apply the method to a subset of the Palomar-Green (PG) quasar sample \citep{sg83}, for which previous measurements of the BLR size are available \citep{kas00}. Summary follows in \S5.

\begin{figure}
\plotone{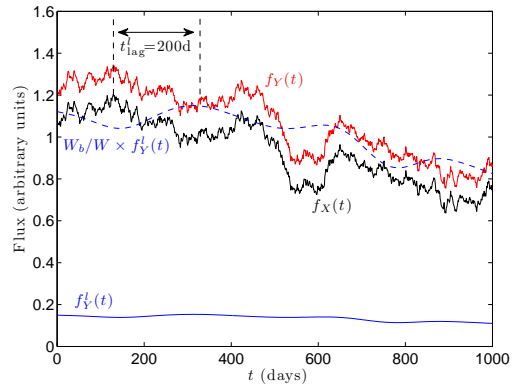}
\caption{A realization of the broadband photometric light-curves of a typical quasar (see text). Flux variations in the $X$ (black) and $Y$ (red) bands show an overall similarity due to the large contribution of highly correlated continuum emission processes to both bands. The response of an emission line to continuum variations in the $X$-band is shown in dashed blue lines ($t_{\rm lag}^l=200$\,days was assumed). The $Y$-band light-curve includes a $13\%$ contribution to the total flux from a broad emission line (solid blue line). A case where the noise level is negligible ($\delta /\left < f \right > =10^{-3}$) was assumed for clarity.}
\label{lc}
\end{figure}

\section{The formalism}

Here we present the algorithm for reverberation mapping the broad emission line regions in quasars using broadband photometric means. We first describe our model for quasar light curves (\S2.1) and then analyze them using new statistical methods that are developed in \S2.2.

\subsection{A model for the light-curve of quasars}

To demonstrate the feasibility of photometric reverberation mapping, we first resort to simulations. Specifically, we rely on current knowledge of the typical spectrum of quasars, and the variability properties of such objects, to construct a model for the light curves in different spectral bands.

\subsubsection{Continuum emission}

Quasar light-curves are reminiscent of red noise, following a power spectrum $P(\omega) \sim \omega^{\alpha}$ ($\omega$ is the angular frequency) with $\alpha\simeq -2$ \citep[and references therein]{giv99}. Here we model the pure continuum light-curve in some band $X$ as a sum of discrete Fourier components, each with a random phase $\phi_i$:
\begin{equation}
f_X^c(t)=\sum_{i=1}^\infty A(\omega_i){\rm sin}(\omega_it+\phi_i).
\end{equation}
The amplitude of each Fourier mode, $A(\omega_i)\propto \omega_i^{\alpha/2}$ and $t$ is time\footnote{We note that, realistically, the power at each frequency is distributed around the above-defined powerlaw, forming an effective envelope with some characteristic width. We ignore this additional complication here, which introduces yet another degree of freedom in the model but does not qualitatively change the results. Real light curves are analyzed in \S4.}. We define the normalized variability measure, $\chi=\sqrt{\sigma_f^2-\delta^2}/\left < f \right >$ [c.f. \citet{kas00} who express this in per-cent], where $ \left < f \right >$ is the mean flux of the light-curve, $\sigma_f$ is its standard deviation, and $\delta$ is the mean measurement uncertainty of the light curve. For the Palomar-Green (PG) sample of quasars \citep{sg83}, the continuum normalized variability measure $\chi_c \sim 0.1-0.2$ \citep{kas00} and, typically, $\delta\gtrsim 0.01$ (see table 1). A light-curve realization, with $\delta\ll \sigma_f$ and $\chi_c\sim0.2$, is shown in Fig. \ref{lc}. 

Quasar emission occurs over a broad range of photon energies, from radio to hard X-ray wavelengths and originates from spatially distinct regions in the quasar. In the optical bands, there is some evidence for red continua to lag behind blue continua \citep{kro91,col98,ser05,bac09}. Without loss of generality, we associate the $Y$-band with the redder parts of the spectrum so that its instantaneous continuum flux may be expressed as a function of the flux in the $X$-band, 
\begin{equation}
f_Y^c(t)= f_X^c * \psi^c \equiv \int_{-\infty}^\infty d\tau f_X^c(\tau) \psi^c(t-\tau),
\label{conVol}
\end{equation}
where $\psi^c(\delta t)$ is the continuum transfer function\footnote{All transfer functions, $\psi$, satisfy the normalization condition $\int_{-\infty}^{+\infty} d(\delta t) \psi(\delta t)=1$.}. Presently, $\psi^c$ is poorly constrained by observations and only the inter-band continuum time delay, $t_{\rm lag}^c$, is loosely determined  \citep[and references therein]{bac09}. Nevertheless, so long as $R_{\rm BLR}/c \gg t_{\rm lag}^c$,  the particular form of $\psi^c$ is immaterial to our discussion (see \S3.2.3). For simplicity, we shall therefore take $\psi^c$ to be a gaussian with a mean of $t_{\rm lag}^c$ and a standard deviation $t_{\rm lag}^c/2$. 

\subsubsection{Line emission}

\begin{figure}
\plotone{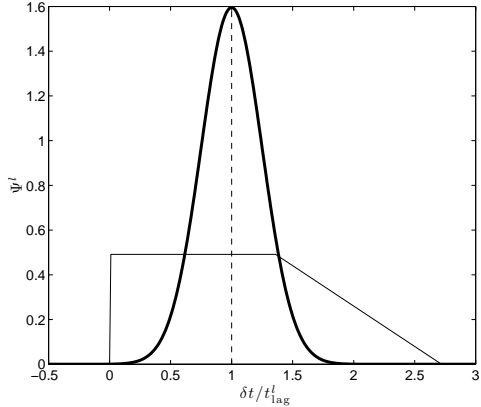}
\caption{The emission line transfer functions used in this work to model the broad lines' response to continuum variations (see text). While a gaussian transfer function (thick line) is relatively well localized in the time domain, the transfer function which is reminiscent of those discussed in \citet{mao90} extends to much longer times. The latter transfer function naturally results in long-range correlations between line and continuum processes (whose effects are discussed in \S3.2.5), and roughly characterizes the reaction of a geometrically thick BLR that is isotropically encompassing the central continuum source. We note that both transfer functions are normalized such that the line-to-continuum time-delay is $\simeq t_{\rm lag}^l$.}
\label{trans}
\end{figure}

Emission by photo-ionized (BLR) gas is naturally sensitive to the continuum level of the ionizing source. In particular, for typical BLR densities and distances from the SMBH, the response time of the photoionized gas to continuum variations is much shorter than the light travel time across the region, and so may be considered instantaneous. Therefore,  the flux in some line $i$, which is emitted by the BLR, can be written, in the linear approximation, as
\begin{equation}
f_i^l= f_X^c * \psi_i^l,
\end{equation}
where the transfer function $\psi_i^l$ depends, essentially, on the geometry of the BLR region as seen from the direction of the observer \citep[and references therein]{hor04}. Our present knowledge of $\psi^l(\delta t)$ for the quasar population as a whole, is, however, rather limited [see \citet{go93} for a few relevant forms for $\psi^l$].  In what follows we shall consider two forms for the transfer function: a) a gaussian kernel with a mean $t_{\rm lag}^l$ and a standard deviation $t_{\rm lag}^l/4$, and b) a transfer function with a flat core in the range $ 0 \leq \delta t \leq 1.36 t_{\rm lag}^l$, with a following linear decline to zero at $\delta t=2.72t_{\rm lag}^l$ (such a transfer function also results in a time-lag being $\simeq t_{\rm lag}^l$). The latter transfer function is motivated by the observations of the Seyfert 1 galaxy NGC\,4151 \citep{mao90} and corresponds to a case where a geometrically thick, spherical shell of broad emission line gas isotropically encompasses the continuum emitting region. The different transfer functions defined above will henceforth be termed (a) the gaussian and (b) the thick-shell BLR models.  The transfer functions are depicted in figure \ref{trans}. The main qualitative difference between the two transfer functions, as far as light curve behavior is concerned,  is that while the gaussian one is relatively localized in time,  the thick shell BLR  model is much broader, and results in longer range correlations between the light curves of the two bands.  Unless otherwise stated, our results are presented for the gaussian model. Cases for which we find a qualitative difference between the two transfer functions are explicitly treated (e.g., \S3.2.5).

\begin{figure*}
\plottwo{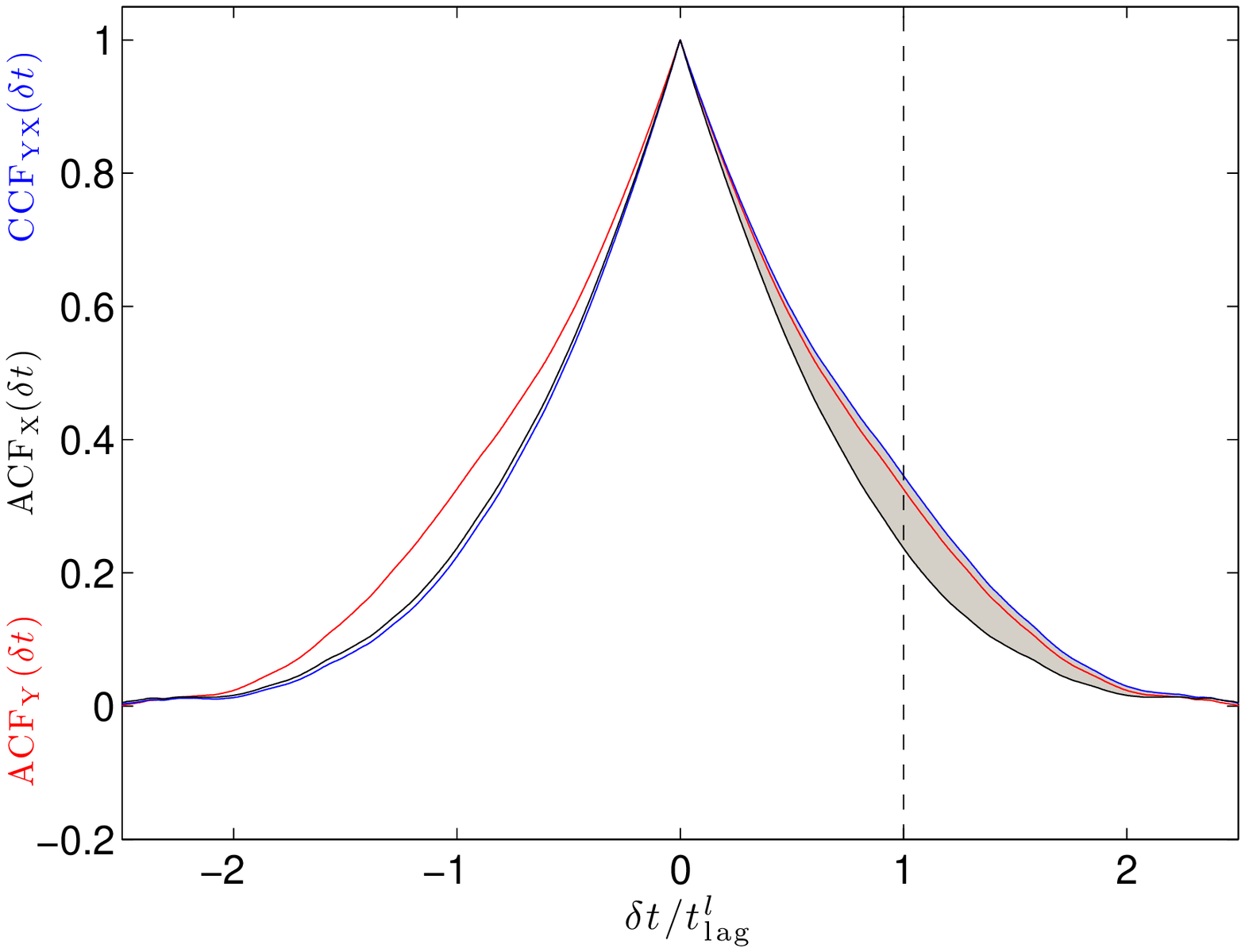}{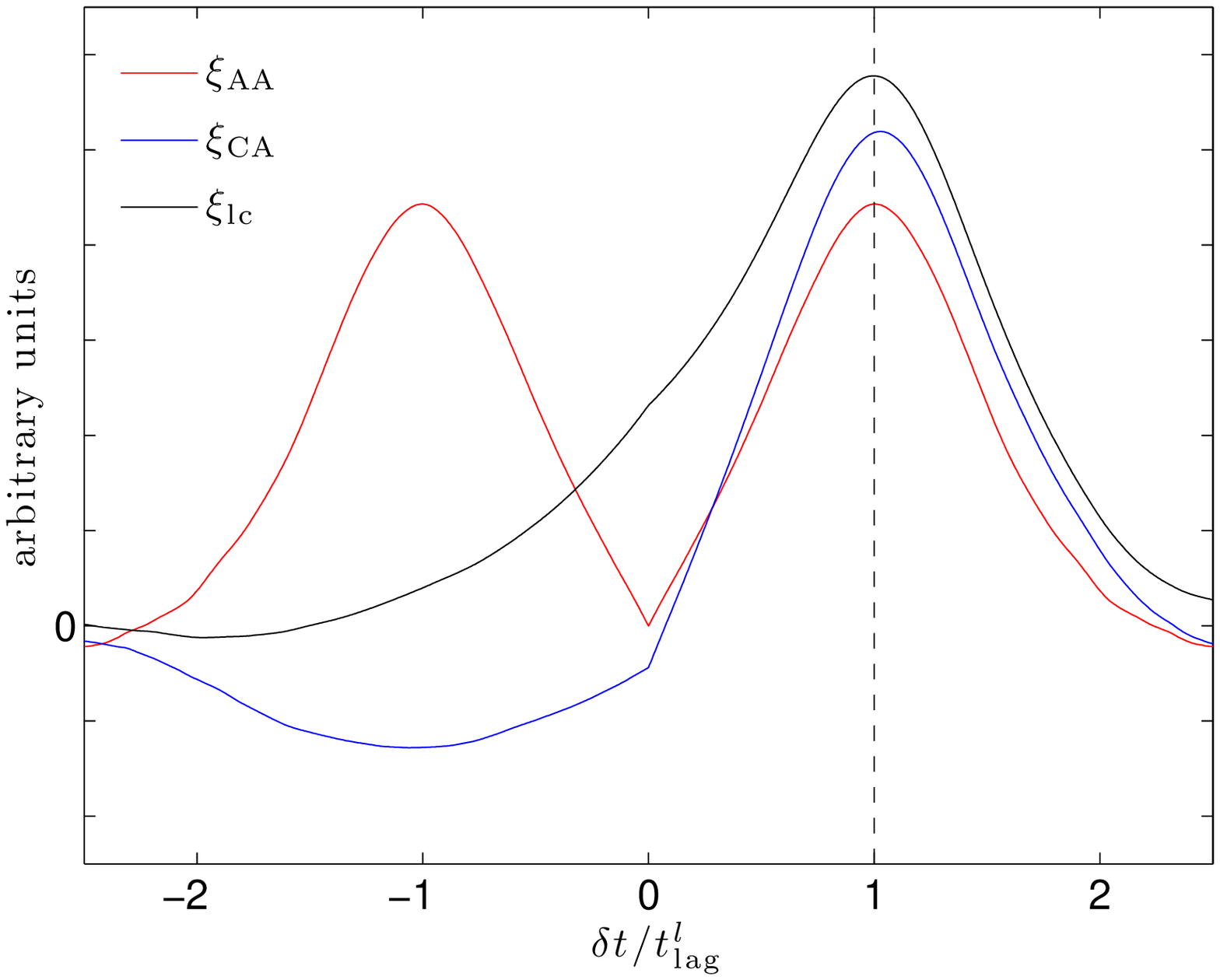}
\caption{Statistical estimators for evaluating the line to continuum time delay. {\it Left:} The ACFs for the $X$ and $Y$ bands are shown (same color coding applies as that in Fig. \ref{lc}), as well as the CCF of the two bands (black curve).  Note the excess power in the ACF of the $Y$-band, as well as in the CCF of the two bands, relative to the ACF of the $X$-band (gray hatched region). This results from relatively long-range correlations between line and continuum variations, which occur on BLR light travel times. The excess power is the sought after signal (see text). {\it Right:} The statistical estimators, $\xi_{\rm CA},~\xi_{\rm AA}$, which quantify the excess power due to line emission, are consistent with $\xi_{lc}$ (see legend). The position of the peaks at positive time-lags (the only ones that preserve causality) is identified with the line-to-continuum time-delay. Evidently, all methods give consistent predictions for the time-lag, which is also in agreement with the input value, $t_{\rm lag}^l$ (dash-dotted black line).}
\label{acfccf}
\end{figure*}

A simulated light-curve of an emission line that is driven by continuum variations (assuming $t_{\rm lag}^l=200$\,days) is shown in figure \ref{lc} (blue curves). Time-delay and smearing effects, with respect to the continuum light-curve, are clearly visible. Here too, we define a normalized variability measure for the line, $\chi_l$, and consider $\chi_l\simeq 0.75 \chi_c$ \citep[see their table 5]{kas00}. The sum of continuum and lines' emission to the total flux in band $Y$ is therefore given by
\begin{equation}
f_Y(t)= f_Y^c(t)+f_Y^l(t) = f_Y^c(t)+\sum_i f_i^l(t),
\label{y}
\end{equation}
where $f_Y^l$ is the total flux in the emission lines. While a similar expression applies, in principle, also for the total flux in band $X$, unless otherwise stated, we shall assume that $f_X=f_X^c$, i.e., that emission line contribution to band $X$ is  either negligible or that the line variability associated with it operates on timescales that are much too short to be resolved by the cadence of the experiment, or too long to be detectable in the time-series. For simplicity, we shall limit our discussion to a single emission line contributing to the $Y$-band (this restriction is relaxed in \S3.2.4).

The contribution of the emission line(s) to the total flux in the $Y$-band requires knowledge of the equivalent width of the broad emission line, $W$ [typically $\lesssim 100$\AA\ in the optical band for strong lines \citep{dvb01}], the transmission curve of the broadband filter, as well as the redshift of the quasar.  For example, for filters with sharp features in their response function, even small redshift changes could lead to substantial differences in the relative contribution of the emission line to the total flux in the band, $\eta$. For a flat filter response function over a wavelength range $W_b$ (typically, $W_b\sim10^3$\,\AA\ for broadband filters), $\eta=W/W_b$. (A more accurate treatment of the line contribution to the broadband flux for the case of Johnson-Cousins filters is given in \S4.) Example light-curves for the $X$ and $Y$ bands, assuming $\eta=0.133$, are shown in figure \ref{lc}. In the following calculations, we shall work with normalized light-curves that have a zero mean and a unit variance, i.e., $f \to (f- \left < f \right > )/\sigma_f$. 

\subsection{Measuring the Line to Continuum Time Delay}

The spectroscopic approach to measuring the size of the BLR cross-correlates pure line and continuum light-curves so that the position of the peak (or the center of mass) of the cross-correlation function (CCF)\footnote{Evaluation of the cross- and auto- correlation functions is done here via the definition of \citet[see his equation 6]{wel99} for the local cross-correlation function. Unless otherwise stated, we use the interpolated CCF method of \citet{gas86}.}, $\xi_{lc}(\delta t)= f_Y^c * f_Y^l$, marks the line to continuum time-delay \citep[and references therein]{pet88}. However, line and continuum light curves cannot be disentangled using pure broadband photometric means and the data include only their combined signal. In fact, light-curves in different bands are rather similar due to the dominant contribution of highly correlated continuum emission processes (Fig. \ref{lc}). Therefore, a naive analysis in which the CCF between the light curves of different bands is calculated would reflect  more on $t_{\rm lag}^c$ than on $t_{\rm lag}^l$ (see Fig. \ref{acfccf}).

Seeking to measure the time-delay between line and continuum processes using broadband photometric means, it is important to realize that instead of spectrally distinguishing between line and continuum light curves, it is possible, under certain circumstances, to separate these processes at the light curve level. In particular, the line light curve,
 $f_Y^l = f_Y-f_Y^c\simeq f_Y-f_X^c$ since $f_Y^c\simeq f_X^c$\footnote{This is expected to be true at least for adjacent bands since the signal is largely due to continuum emission processes, such as black-body radiation, which cover a broad wavelength range. Interestingly, this broad similarity of light curves in different bands seems to be true also for the (hidden) far-UV and optical bands given the success of the spectroscopic reverberation mapping technique in measuring line to continuum time-delays. The effects of inter-band time-delay and time-smearing of the continuum signal may account for the observed structure function differences between adjacent bands \citep{mcl10}, and are further discussed in \S3.2.3.}. This is justified for quasars since the continuum auto-correlation timescale [$\gtrsim 10^2$\,days \citep{giv99}, possibly reflecting on viscous timescales in the accretion flow] is much longer than $t_{\rm lag}^c$ [being of order days \citep{ser05} and resulting from irradiation processes in the accretion disk, which take place on light travel time scales]. Furthermore, as $f_X=f_X^c$, $f_Y^l\simeq f_Y-f_X$, and we obtain,
\begin{equation}
\xi_{lc}\simeq \xi_{\rm CA}(X,Y) \equiv (f_Y-f_X) *f_X=f_Y * f_X -f_X * f_X,
\label{theta}
\end{equation}
where the rightmost term is just the difference between the CCF of the light-curves and the auto-correlation function (ACF) of the light curve with the negligible line contribution (see \S3.2 for cases in which emission lines contribute also to $f_X$). Graphically, this difference is simply the shaded region shown in the left panel of figure \ref{acfccf}, which arises due to correlated line and continuum processes adding power on $t_{\rm lag}^l$-scales. Using these statistics, and given a rather non-restrictive set of assumptions (see more in \S3), it seems, in principle, possible to recover the line-to-continuum time-delay using {\it pure} broadband photometric data. In particular, for the example shown in figure \ref{acfccf}, $\xi_{\rm CA}$ peaks at the correct time-lag, and is in agreement with the time-lag estimate as obtained using the spectroscopic method, i.e., using $\xi_{lc}$. 

The estimator defined by equation \ref{theta} provides a good approximation for $\xi_{lc}$ in cases where the emission line contribution to the band is sub-dominant (see below). For the more general case (e.g., when narrow band filters are concerned), it is possible to define a somewhat more generalized version where $\xi_{\rm CA}'\equiv f_Y * f_X - (1-\eta) f_X * f_X$, which is applicable for cases in which the emission line contribution to the $Y$-band is considerable or even dominant\footnote{In this case, $\lim_{\eta \to 1 }\xi_{\rm CA}'=\xi_{lc}$. For $\eta \ll 1$, as appropriate for broadband photometry, $\xi_{\rm CA}'\simeq \xi_{\rm CA}$.}. Nevertheless, with this definition, $\eta$ needs to be observationally known, which is not always the case, especially when large (photometric) redshift uncertainties are concerned, the filter response function has sharp features \citep{che11}, or the quasar spectrum, if available, is noisy. Focusing on broadband photometric light curves in this work, the contribution of emission lines to the band is, typically, of order a few per cent (see \S4). We therefore restrict the following discussion to $\xi_{\rm CA}$ as defined by equation \ref{theta}.

\begin{figure*}
\epsscale{1.2}
\plotone{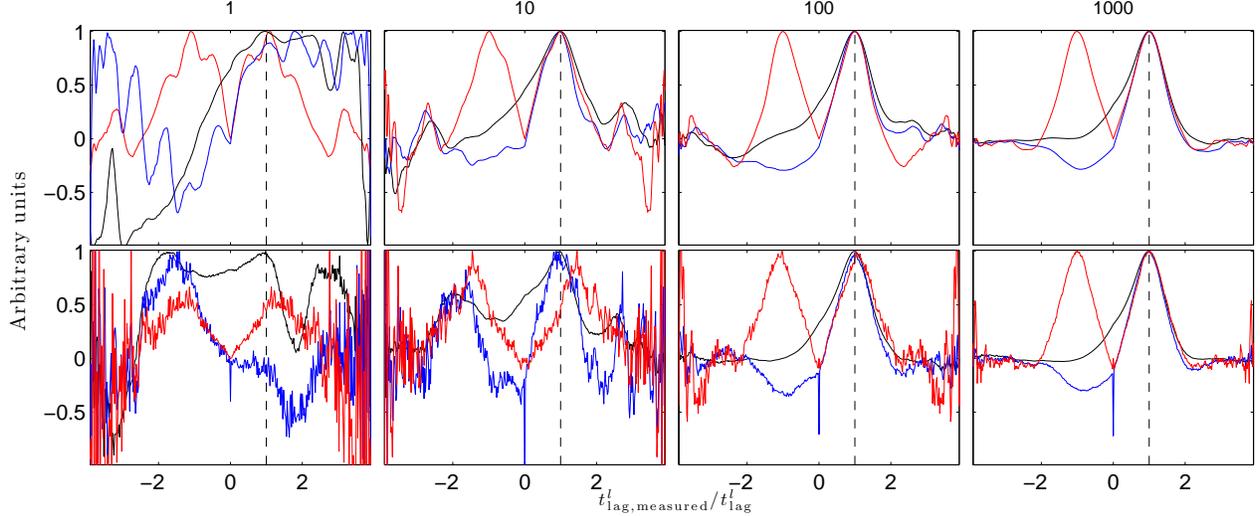}
\caption{Averages of $\xi_{\rm CA}$ (blue) ,$\xi_{\rm AA}$ (red) as well as for $\xi_{lc}$ (black) for ensembles of statistically equivalent quasars (the number of objects in each ensemble is indicated above each column). Upper (lower) panels correspond to a noise-level $\delta / \left < f \right >=0$ (0.03). In all simulations the total duration of the light curve is $4t_{\rm lag}^l$, and visits are $t_{\rm lag}^l/50$ apart. The relative contribution of the emission line to the filter flux, $\eta \simeq 0.13$. Clearly, finite sampling of the light curves, even in the absence of measurement noise, could result in the peaks being offset with respect to $t_{\rm lag}^l$. This is true for individual objects as well as for small samples.  Generally, however, upon averaging over many objects, the mean estimators peak at $\delta t \simeq t_{\rm lag}^l$ (right panels). When measurement noise is present, larger ensemble are required for the peak of the $\left < \xi_{\rm CA} \right > ,~ \left < \xi_{\rm AA} \right >$ and $\left < \xi_{lc} \right >$ to coincide with $t_{\rm lag}^l$ to good accuracy (e.g., compare the top and bottom panels in the second column from the left). Overall, the photometric and the spectroscopic estimators give consistent results, which is especially true for large ensembles.}
\label{ensembles}
\end{figure*}

It is possible to define an additional statistical estimator to measure line to continuum time-delays using broadband photometry. Specifically, by using the fact that, for broadband filters and to zeroth order approximation, $f_X \simeq f_Y$ (see Fig. \ref{lc}), we may approximate $\xi_{\rm CA}$ by 
\begin{equation}
\xi_{\rm AA}(X,Y) \equiv f_Y * f_Y -f_X * f_X.
\label{xiaa}
\end{equation}
This estimator is just the difference between the ACFs of the light-curves in each band. A more intuitive way to understand the above definition may be gained by looking at figure \ref{acfccf}, where the ACF of filter $Y$ shows more power on $t_{\rm lag}^l$ timescales due to relatively long-term correlations between line and continuum processes. In contrast, the ACF of filter $X$, being devoid of line emission, does not show a corresponding power-excess. We note that both definitions of $\xi$ are quasi anti-symmetric with respect to $X$ and $Y$ exchange.

We find that, as far as $t_{\rm lag}^l$ measurements are concerned, both estimators, $\xi_{\rm CA}$ and $\xi_{\rm AA}$, give, in principle, consistent results (see Fig. \ref{acfccf}). While $\xi_{\rm CA}$ has the advantage of being a more reliable estimator of $\xi_{lc}$ (see more in \S3), $\xi_{\rm AA}$ has the advantage of not containing a cross-correlation term of the form $f_Y * f_X$. This property allows one to evaluate $\xi_{\rm AA}$ even in cases where  non-contemporaneous light curves are available so long as a) both light curves are adequately sampling the line transfer function, and b) quasar variability results from a stationary process (see \S3.2.5). Moreover, if the variability properties for an ensemble of objects (of some given property) are statistically similar, and the auto-correlation terms, $f_Y * f_Y$ and  $f_X * f_X$, are well determined in the statistical sense (see  \S3.1.1), then it is not required for the same objects to be observed in both bands. This opens up the possibility of monitoring different parts of the sky in each band, and using the acquired data to statistically measure the mean line to continuum time-delay. That being said, large ensembles may be required to achieve a reliable measurement of $t_{\rm lag}^l$ since $\xi_{\rm AA}$ is, in fact,  an approximation to an approximation of $\xi_{lc}$, and its results are expected to be somewhat less reliable than both $\xi_{\rm CA}$ and $\xi_{lc}$ (this is further discussed in \S3). 

A note is in order concerning the maximal value of $\xi_{\rm AA}(\delta t)$ and $\xi_{\rm CA}(\delta t)$. Unlike the spectroscopic technique where $\xi_{lc}(\delta t)$ is, in fact, the Pearson correlation coefficient and its  value measures how well continuum and line processes are correlated at any $\delta t$, for the photometric estimators defined here, the peak value depends also on the relative contribution of the emission line(s) to the broad band flux. The latter depends not only on the properties of the quasar, but also on the transmission curve of the broadband filter, the variability measure of the line, and the redshift of the object. Therefore, the significance of the result is not directly implied by the value of the peak in $\xi_{\rm AA}$ and $\xi_{\rm CA}$, and quantifying it requires a different approach, as we shall discuss in \S\S3,4.

\section{Simulations}

While the reliability of the spectroscopic reverberation mapping technique has been assessed in numerous works [e.g., \citet{gas87} and references therein], we have yet to demonstrate it for the photometric method. In this section we take a purely theoretical approach and use a suite of Monte Carlo simulations to test the method and quantify its strengths and weaknesses. 

\subsection{Sampling \& Measurement Noise}

All light curves are finite with respect to their duration time, $t_{\rm tot}$, as well as the sampling interval, $t_{\rm sam}$. This fact alone introduces potentially substantial "noise" in the cross-correlation analyses (whether photometric or spectroscopic). The effect of finite sampling is seen in Fig. \ref{ensembles} (left panels) where the light curves of individual objects are analyzed, showing that $\xi_{\rm CA},~\xi_{\rm AA}$ formally peak at values that could be significantly removed from $t_{\rm lag}^l$. This implies that even in the ideal case, where measurement noise is negligible, the deduced time-lag might be significantly offset from the true value. While similar situations occur also with spectroscopic data, our simulations indicate that the photometric estimators are, statistically, more prone to such effects.

In addition to finite sampling there is measurement noise, which reflects on the ability of any statistical estimator to recover the true time-lag with good accuracy. On an individual case basis, the effect of noise is seen to give rise to more erratic $\xi_{\rm CA}$ and $\xi_{\rm AA}$ behavior, potentially leading to substantial deviations in the position of the peak from the true time-lag. In fact, for the particular case shown (Fig. \ref{ensembles}), $\xi_{\rm CA}$ has a kink at $t_{\rm lag}^l$ and peaks only at $\sim 3t_{\rm lag}^l$. In this particular case, $\xi_{\rm AA}$ happens to be less affected by noise, and formally peaks at $\sim 1.15t_{\rm lag}^l$ due to a narrow feature being the result of noise.

\begin{figure}
\plotone{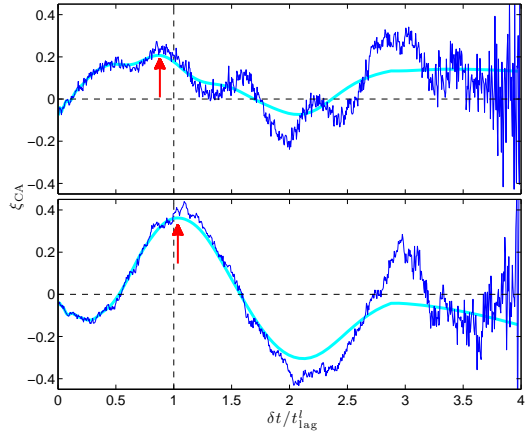}
\caption{Photometric time-lag measurements for individual quasars having noisy data ($\delta/\left < f \right >=0.03/0.01$ in the top/bottom panel). Calculated $\xi_{\rm CA}$'s  (blue curves) show multiple peaks, some of which are narrow and peak at $\delta t > t_{\rm lag}^l$, potentially with large amplitudes (note the feature at $\delta t/t_{\rm lag}^l\simeq 3$).  Smoothing of the signal using a running mean algorithm over a uniform logarithmic grid suppresses narrow features and highlights the peak associated with the time lag (cyan curves; see text). Red arrows mark the peak at $t_{\rm lag,measured}^l$, which, in this example, deviates only slightly from the input lag (dashed vertical line at $\delta/t_{\rm lag}^l=1$).}
\label{smooth}
\end{figure}

Sampling and measurement noise are manifested as small scale fluctuations in all statistical estimators. In particular, in some cases (see Figs. \ref{ensembles} \& \ref{smooth}), large amplitude fluctuations are observed, which peak at long timescales, $\delta t / t_{\rm lag}^l>1$, and their value could  exceed the value of the true peak (i.e., that which is associated with, and peaks at, $\sim t_{\rm lag}^l$). Clearly, an algorithm that would aid to properly identify the relevant signal, would be of advantage. To this end we use a physically-motivated algorithm related to the geometry of the BLR: the fact that the BLR encompasses the continuum emitting region means that the width of its line transfer function should be of order the time-lag. Therefore, one expects the relevant signal in $\xi_{\rm CA},~\xi_{\rm AA}$ to peak at $t_{\rm lag}^l$ with a width, $\Delta (\delta t)\sim t_{\rm lag}^l$. In particular, narrower features with $\Delta (\delta t)/t_{\rm lag}^l \ll 1$ are likely to be merely the result of noise or sampling, and although they could reach large amplitudes, they are of little relevance to the time-lag measurement. 

The aforementioned properties of the line transfer function motivate us to consider the following algorithm to aid in the identification of the time-lag: we resample $\xi_{\rm CA}$ and $\xi_{\rm AA}$ onto a uniform logarithmic grid, which preserves resolution in $\Delta (\delta t)/\delta t$ units. We then apply a running mean algorithm such that the boxcar size corresponds to a fixed range in logarithmic scale. This procedure smooths out narrow features having $\Delta (\delta t)/\delta t \ll 1$. Examples for smoothed statistical estimators are shown for two levels of noise in figure \ref{smooth}. Clearly, narrow features at large $\delta t$ are significantly suppressed without diminishing the power of similar $\Delta (\delta t)$ features centered at smaller $\delta t$. It should be noted, however, that using this technique does not assure that the correct time-lag is recovered (see above). Furthermore, in some cases, multiple peaks may be found and additional methods are required to assess their significance (see \S4). 

\begin{figure}
\plotone{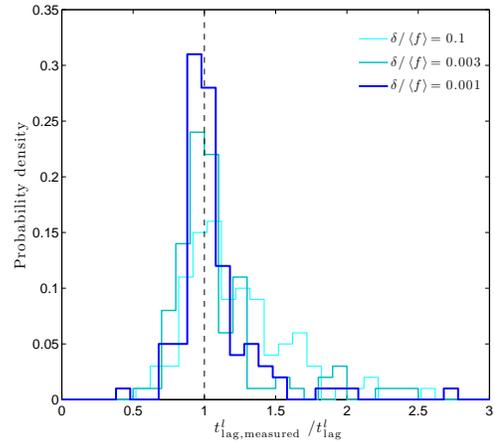}
\caption{Measured time-lag distributions for a sample of 100 quasars (having 800 points in each light curve and with $t_{\rm tot}/t_{\rm lag}^l=5$; $\eta\simeq 0.13$ was assumed), and for several noise levels (see legend). Clearly, for all cases, the distributions peak at the input time lag (dashed line), demonstrating the feasibility of the photometric approach for time-lag measurement. As expected, noisier light curves result in broader time lag distributions. The distributions show a tail extending to large $t_{\rm lag,measured}^l/t_{\rm lag}^l$-values and are clearly non-gaussian.}
\label{smooth1}
\end{figure}

Testing for the accuracy of time-lag measurements using noisy data, we have conducted uniformly sampled light curves simulations of a quasar (100 realizations were used having 800 points in each light curve, with a total light curve duration, $t_{\rm tot}$, satisfying $t_{\rm tot}/t_{\rm lag}^l=5$ and $\eta\sim 0.13$) at several levels of measurement noise. For each realization, $\xi_{\rm CA}$ was calculated and $\delta t$ where it peaks was identified with $t_{\rm lag,measured}^l$ and logged. The resulting time-lag statistics are shown in figure \ref{smooth1}. The distribution is clearly non-gaussian and shows a peak at the input time lag with a tail extending to long time-lags. Not unexpectedly, the dispersion in time-lag measurements increases with the noise level. For the particular examples shown, taking $\delta / \left < f \right >=0.01$ [a noise level typical of photometric light curves of quasars \citep{kas00}], $\sim 60$\% of the measurements will result in $t_{\rm lag, measured}^l$ being within $\pm 10$\% of the true lag. For noisier data with $\delta / \left < f \right >=0.1$, only $\sim 30$\% of all realizations result in a time lag within $\pm10$\% of the true time lag.  These calculations demonstrate that photometric reverberation mapping in the presence of measurement noise typical of photometrically monitored quasars is indeed feasible, and that the measurement is, in principle, reliable. When noisier data are concerned, the uncertainty on the time-lag is larger. 

\begin{figure*}
\plottwo{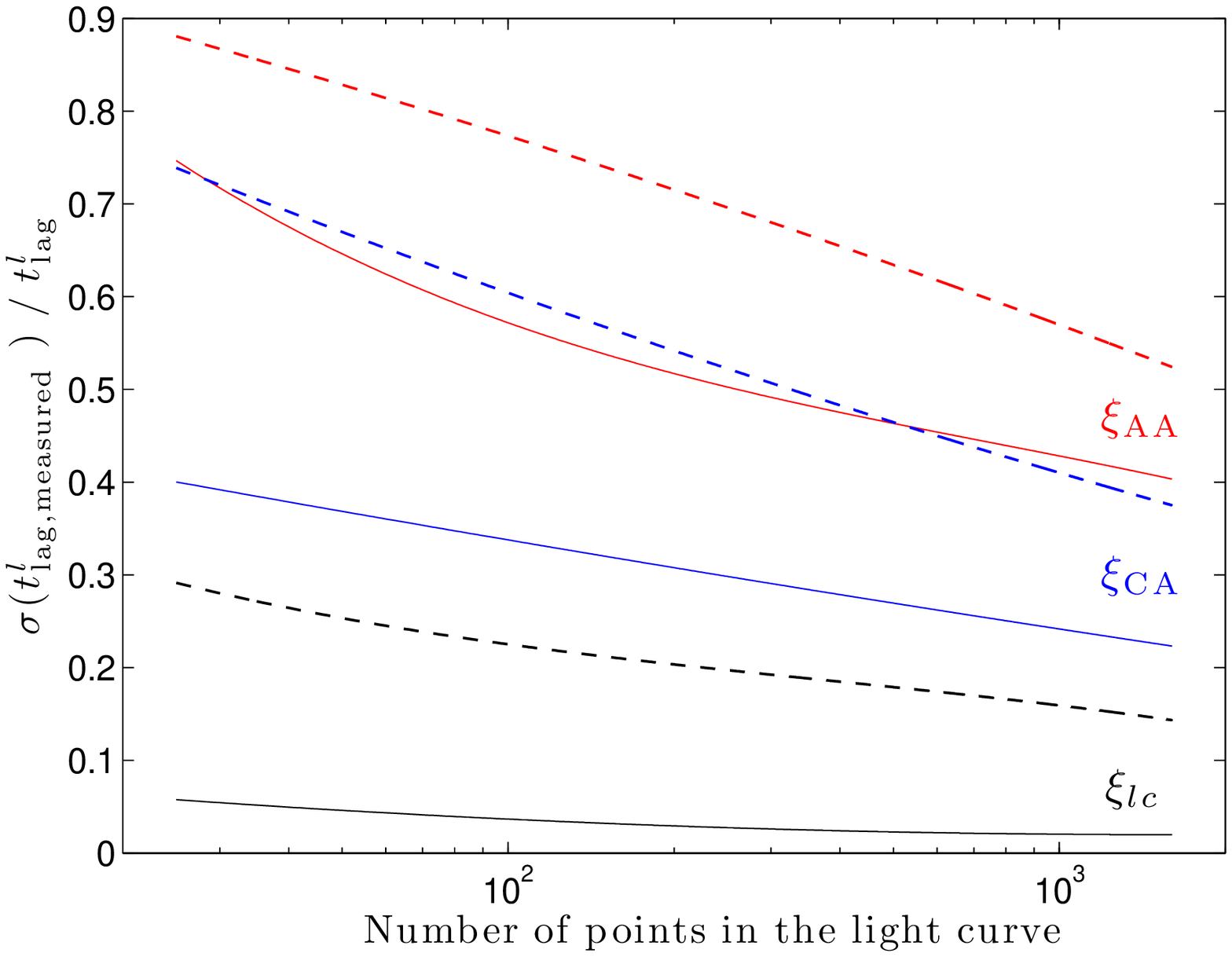}{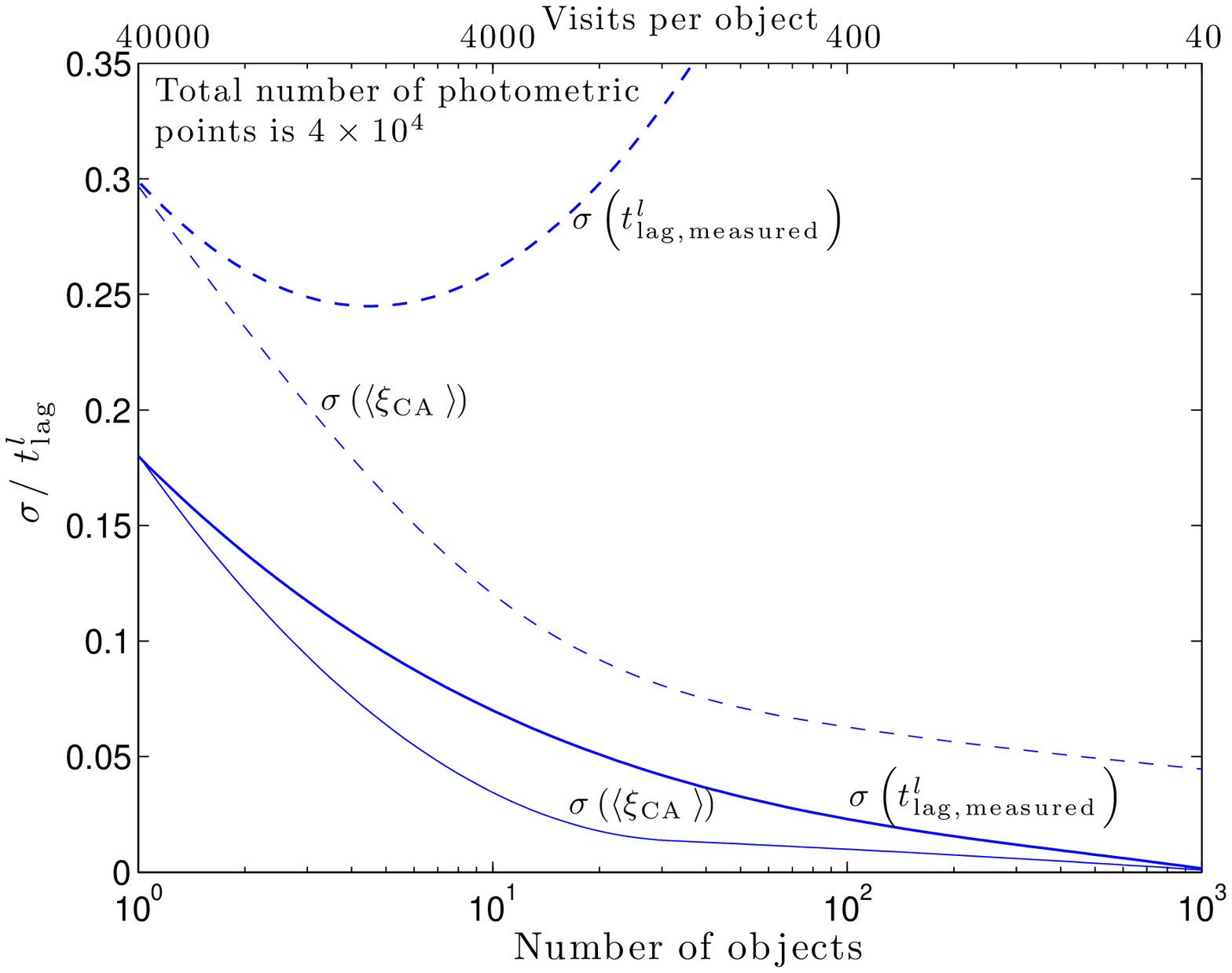}
\caption{Relative measurement uncertainties on $t_{\rm lag,measured}^l$.  {\it Left:} One standard deviation of the  measured time-lag distribution functions (i.e., the uncertainty on $t_{\rm lag,measured}^l$) as obtained by measuring the time delay for individual realizations in a large ensemble of objects (having the same BLR size and sampling characteristics as those of figure \ref{smooth}).  One hundred simulations were used to calculate each data point (6 equally log-spaced points were calculated in each curve and a second degree polynomial fit to the results is shown). Noise level $\delta/\left < f \right >=0.01~(0.1)$ is shown in solid (dashed) lines, and the usual color coding scheme applies ($\xi_{\rm CA}/\xi_{\rm AA}/\xi_{lc}$ results in blue/red/black curves). Clearly, better signal-to-noise is obtained by either reducing the measurement noise or by better sampling the light curve. Generally, $\xi_{\rm CA}$ is the more reliable photometric estimator, while both photometric methods are less accurate than the spectroscopic one ($\xi_{lc}$). {\it Right:} The uncertainty on the {\it mean} time-lag deduced for an ensemble of objects using $\xi_{\rm CA}$, for two noise levels ($\delta/\left < f \right >=0.01,~0.1$ are shown in solid and dashed curves, respectively, and smoothing was applied). The total observing effort (number of points for the entire sample) was held fixed. The time-lag was deduced in two ways: (1) by measuring the time-lag for individual objects with the $\xi_{\rm CA}$ estimator and taking the mean result (thick lines), and by measuring the position of the peak of $\left < \xi_{\rm CA} \right >$ (thin lines). The uncertainty on the mean time-lag reported for the latter quantity was deduced from multiple simulations of statistically equivalent ensembles. The results show that, for the same observing effort, monitoring more objects with lower cadence (so long as the line transfer function is properly sampled) results in more accurate time-lag measurements. Also, measuring the peak in the mean statistical estimator provides more accurate results than measuring the time-lag for individual objects [$\sigma(\left < \xi_{\rm CA} \right >)<\sigma(t_{\rm lag,measured}^l)$].}
\label{1sigma}
\end{figure*}

To better quantify the ability of the photometric method to uncover the time-lag under various observing strategies, we repeated the above analysis for two noise levels ($\delta / \left < f \right >=0.01,0.1$) while varying the number of points in the light curves (i.e., observing visits). More specifically, we maintained a fixed $t_{\rm tot}/t_{\rm lag}^l=5$ and varied the sampling period, $t_{\rm sam}$ (we made sure that $t_{\rm lag}^l/t_{\rm sam} \geq 5$ so that the BLR transfer function is never under-sampled). Monte Carlo simulations were carried out to obtain reliable statistics, and we report the one standard deviations of the measured time-lag distributions, $\sigma (t_{\rm lag,measured}^l)$, in figure \ref{1sigma} (also shown, for comparison, are the corresponding $\xi_{lc}$-statistics). As expected, better sampled light curves result in a lower dispersion in $t_{\rm lag,measured}^l$ for all statistical methods. Also, unsurprisingly, $\xi_{\rm CA}$ is the more reliable photometric estimator for the time-lag, and both photometric estimators are worse than $\xi_{lc}$. In particular, quasar light curves having $\sim 100$ points in each filter with photometric errors of order 1\% result in a $\xi_{\rm CA}$-deduced time-lag dispersion of order $\sim 35\%$. Similar quality light curves can be found in the literature \citep{giv99,kas00}, and are analyzed in \S4.

Consider the following question: having finite resources, is it better to observe fewer objects with better cadence or more objects having sparser light curves? [We assume that the {\it cadence is adequate} for sampling the line transfer function (see more in \S3.2), and that the same signal-to-noise is reached in both cases.] The answer to this question is given in figure \ref{1sigma} (right panel) where the total number of visits was kept fixed yet the number of objects in the ensemble varied (correspondingly, the number of visits per object changed to maintain their product constant). Interestingly, for the same observing effort, it is advantageous to observe more objects with worse cadence.  Using this strategy, one is less sensitive to unfavorable variability patterns of particular quasars. For the case considered here, the time-lag uncertainty drops by a factor $\sim 3$ by increasing the sample size from one to ten objects (the uncertainty for the case of ensembles was determined from Monte Carlo simulations of many statistically similar ensembles). Similar results are obtained also for $\xi_{\rm AA}$ (not shown).

\subsubsection{Ensemble averages}

In the upcoming era of large surveys, statistical information may be gathered for many objects. For example, the LSST is expected to provide photometric light curves for $>10^6$ quasars having a broad range of luminosities and redshifts. With such large sets of data at one's disposal, sub-samples of quasars may be considered having a prescribed set of properties (e.g., luminosity and redshift), and the line to continuum time-lag measurement may be quantified for the ensembles.  In this case, it is possible to measure the characteristic line-to-continuum time-delay in two ways: by measuring the peak of $\xi_{\rm CA} ( \delta t )$ or $\xi_{\rm AA} ( \delta t )$ on a case by case basis and then constructing time-lag distributions to measure their moments, or by defining the ensemble averages of the statistical estimators. Specifically, we define $\left < \xi_{\rm CA} (\delta t) \right >,~ \left < \xi_{\rm AA} (\delta t) \right >$ as the arithmetic averages\footnote{We similarly define, for comparison purposes, $\left < \xi_{lc} (\delta t) \right >$. These definitions were, in fact, previously used to compute the curves shown in figure \ref{acfccf}.} of the corresponding statistical estimators at every $\delta t$. (Alternative definitions for the mean do not seem to be particularly advantageous and are not considered here.)  The average statistical estimators for ensembles of varying sizes are shown in figure \ref{ensembles}. As expected, averaging leads to a more robust measurement of the peak, as the various $\xi_{\rm CA} (\delta t)$'s "constructively-interfere" at $\sim t_{\rm lag}^l$. The ensemble size required to reach a time-lag measurement of a given accuracy naturally depends on the noise level.

We find that measuring the time-lag from the peak of $\left < \xi_{\rm CA} \right >$ (or $\left < \xi_{\rm AA} \right >$) is more accurate than measuring the time-lags for individual objects in the ensemble, and then quantifying the moments of the distribution (Fig. \ref{1sigma})\footnote{The uncertainty on the time-lag, as deduced from $\left < \xi_{\rm CA} \right >$, was calculated by simulating numerous statistically-equivalent ensembles of objects, calculating $\left < \xi_{\rm CA} \right >$, and measuring the time-lag from its peak. A time-lag distribution is obtained with the uncertainty given by its standard deviation, $\sigma(\left < \xi_{\rm CA} \right >)$.}. This is particularly important when noisy data are concerned, in which case reliable time-lag measurements for individual objects are difficult if not impossible to obtain. To see this, we note the rising uncertainty on the deduced time-lag, $\sigma(t_{\rm lag,measured}^l)$, as the number of visits per object is reduced (and despite the fact that a correspondingly larger sample of objects is considered; see the right panel of Fig . \ref{1sigma} for a case in which $\delta/\left < f \right >=0.1$). In contrast, when working with $\left < \xi_{\rm CA} \right >$-statistics, the uncertainty monotonically declines so long as the sampling of the transfer function is adequate. 

The above analysis highlights the importance of using large samples of objects (even with less than ideal sampling) in addition to studying individual objects having relatively well-sampled light curve.  In particular,  by using large enough samples of quasars, the effects of noise, and even sparse sampling, may be overcome. 

\begin{figure}
\plotone{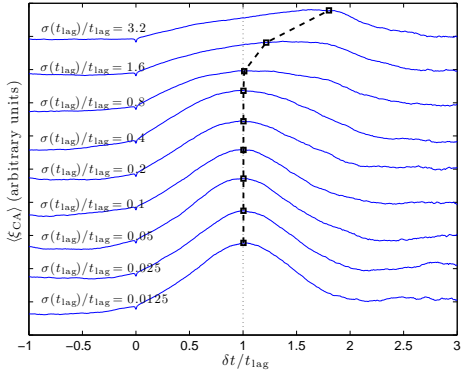}
\caption{The effect of scatter in the broad line region properties on (arbitrarily scaled) $\left < \xi_{\rm CA} (\delta t) \right >$. A large ensemble of objects  (typically 100-1000) was used, whose time-lag distribution follows a modified Gaussian (see text), which is characterized by the mean time-lag, $t_{\rm lag}$ and its standard deviation, $\sigma(t_{\rm lag})$. $t_{\rm tot}/t_{\rm lag}=4$ was assumed. For $\sigma(t_{\rm lag})/t_{\rm lag}<1$, the distribution is nearly Gaussian, $\left < \xi_{\rm CA} (\delta t) \right >$ is, likewise, symmetric, and its peak is at $t_{\rm lag}$. For $\sigma(t_{\rm lag})/t_{\rm lag}\gtrsim 1$, the time-lag distribution is, by definition, modified (see text), and the effective mean time lag is $>t_{\rm lag}$ (black squares for each curve). This, and the fact that, for some objects, $t_{\rm lag}^l \sim t_{\rm tot}$ (hence the line transfer function is poorly sampled) result in an asymmetric $\left < \xi_{\rm CA} (\delta t) \right >$, which only roughly peaks at the effective mean time lag. }
\label{lrs}
\end{figure}

Current studies suggest that there is a scatter of 30\%-40\% in the BLR size vs. quasar luminosity relation \citep{ben09,kas00}.  To assess the effect of such a scatter on $\left < \xi_{\rm CA} \right >$ (similar results apply also for $\left < \xi_{\rm AA} \right >$), we averaged the results for an ensemble of simulated objects whose time-lags follow a Gaussian distribution with a mean $t_{\rm lag}$ and standard deviation $\sigma(t_{\rm lag})$. We modify this Gaussian such that the input lag is always positive definite: for  $\sigma(t_{\rm lag})/t_{\rm lag} \ll 1$ the distribution is Gaussian while for $\sigma(t_{\rm lag})/t_{\rm lag} \sim 1$ there is an excess of systems with zero lag. $\left < \xi_{\rm CA} (\delta t) \right >$, is shown in figure \ref{lrs}: evidently, the average statistical estimator peaks at the mean of the distribution so long as $\sigma(t_{\rm lag})/t_{\rm lag} < 0.8$. For larger ratios, the distribution considerably deviates from Gaussian and its mean is $>t_{\rm lag}$. In this case, $\left < \xi_{\rm CA} (\delta t) \right >$ still peaks around the mean but its shape is highly asymmetric. This is partly due to the time-lag distribution, but also so due the fact that the transfer functions with the largest $t_{\rm lag}^l$'s are not adequately sampled by the light curve ($t_{\rm tot}/t_{\rm lag}^l=4$ was assumed). To conclude, photometric reverberation should yield accurate mean BLR sizes for a sample of quasars, given the observed scatter in the BLR properties.

\subsection{Systematics}

We identified several potentially important biases in the method, which we shall now quantify using a suite of Monte Carlo simulations. We work in the limit of large ensembles so that the effect of sampling and measurement noise are negligible (typically, the results of several hundreds of realizations are averaged over and the line to continuum time delay is identified with the peak of $\left < \xi_{\rm CA} (\delta t) \right >$ and $\left < \xi_{\rm AA} (\delta t) \right >$). 

\subsubsection{Measurement noise}

\begin{figure}
\plotone{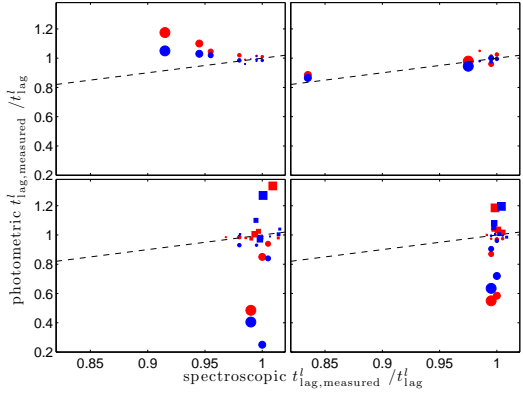}
\caption{Biases in measuring line to continuum time delays as calculated for the spectroscopic (abscissa) and photometric methods (ordinate). Dashed line is for a 1:1 ratio. Upper-left panel shows the effect of noise with the largest data point [blue (red) reflecting on $\xi_{\rm CA}$-($\xi_{\rm AA}$-) deduced measurements] corresponding to $\delta/\left < f \right >=0.512$ with subsequently smaller size symbols corresponding to values smaller by a factor 2 (the same symbol-size value applies to all panels). Upper-right panel shows the bias as a function of $t_{\rm gap}$ and implies consistency between the spectroscopic and photometric results (largest symbol is for $t_{\rm gap}/t_{\rm lag}^l=2$). Lower-left panel quantifies the bias for the $X$-band leading (circles) or trailing (squares) the $Y$-band. Largest symbol is for $\vert t_{\rm lag}^c \vert /t_{\rm lag}^l\simeq 0.3$. Lower-right panel shows the bias in the presence of a subordinate emission line whose time-lag is $t_{\rm lag}^l/2$, which is contributing to the $Y$ (circles) and to the $X$ (squares) filter. The largest symbol corresponds to $\eta_s/\eta = 0.8$ (see text).}
\label{biases}
\end{figure}

Biases are present in the time-lag measurement from noisy data. These are negligible for noise levels $\delta / \left < f \right > < \eta$ but may surface otherwise. Specifically, for a (Gaussian) noise level of $\sim 50\%$, and a relative line contribution to the band of 10\%, both photometric estimators over-estimate $t_{\rm lag}^l$ by as much as 20\% with $\xi_{\rm AA}$ being more biased than $\xi_{\rm CA}$ (Fig. \ref{biases}). This bias could  therefore  be present in  situations where large and noisy ensembles are considered, or when weak emission lines are targeted. Quantifying the bias requires good understanding of the noise properties.

A related bias exists when the noise level in the two bands differs substantially. More specifically, when $\vert \delta_X/\left < f_X\right>-\delta_Y/\left < f_Y\right> \vert \sim \eta$ ($\delta_X/\delta_Y$ is the noise level in the $X$-/$Y$- bands), our simulations show that $t_{\rm lag,measured}$ may differ from $t_{\rm lag}$ by as much as 25\%. Whether an over-estimate or an under-estimate of $t_{\rm lag}$ is obtained depends on the particular noise properties, as mentioned above. Observationally, one should therefore aim to work with data having comparable noise levels for the different bands to obtain a bias-free result. As noted above, for relatively strong lines or low noise levels, such biases are negligible.

Varying seeing conditions leading to aperture losses, as well as improperly accounting for stellar variability in the field of the quasar, could lead to correlated errors between the bands. This in turn could lead to artificial zero-lag peaks in the correlation function of the light curves \citep{wel99}, and several solutions  have been devised to overcome this problem \citep[and references therein]{ale97,zu11}. Nevertheless, the formalism developed here subtracts off, by construction,  the contribution of similar processes (e.g., correlated noise) to both light curves. In fact, as far as the algorithm is concerned, the continuum contribution to the $Y$-band is merely an effective correlated noise to be corrected for, which it does. Therefore, photometric reverberation mapping, as proposed here, is considerably less susceptible to the effects of correlated noise than the spectroscopic method. That being said, if the noise between the bands correlates over sufficiently long (e.g., $\gg$\,day) timescales, a secure detection of the line-to-continuum time delay may be more challenging (for an effectively similar case see \S3.2.3).

\subsubsection{Sampling}

\begin{figure*}
\plottwo{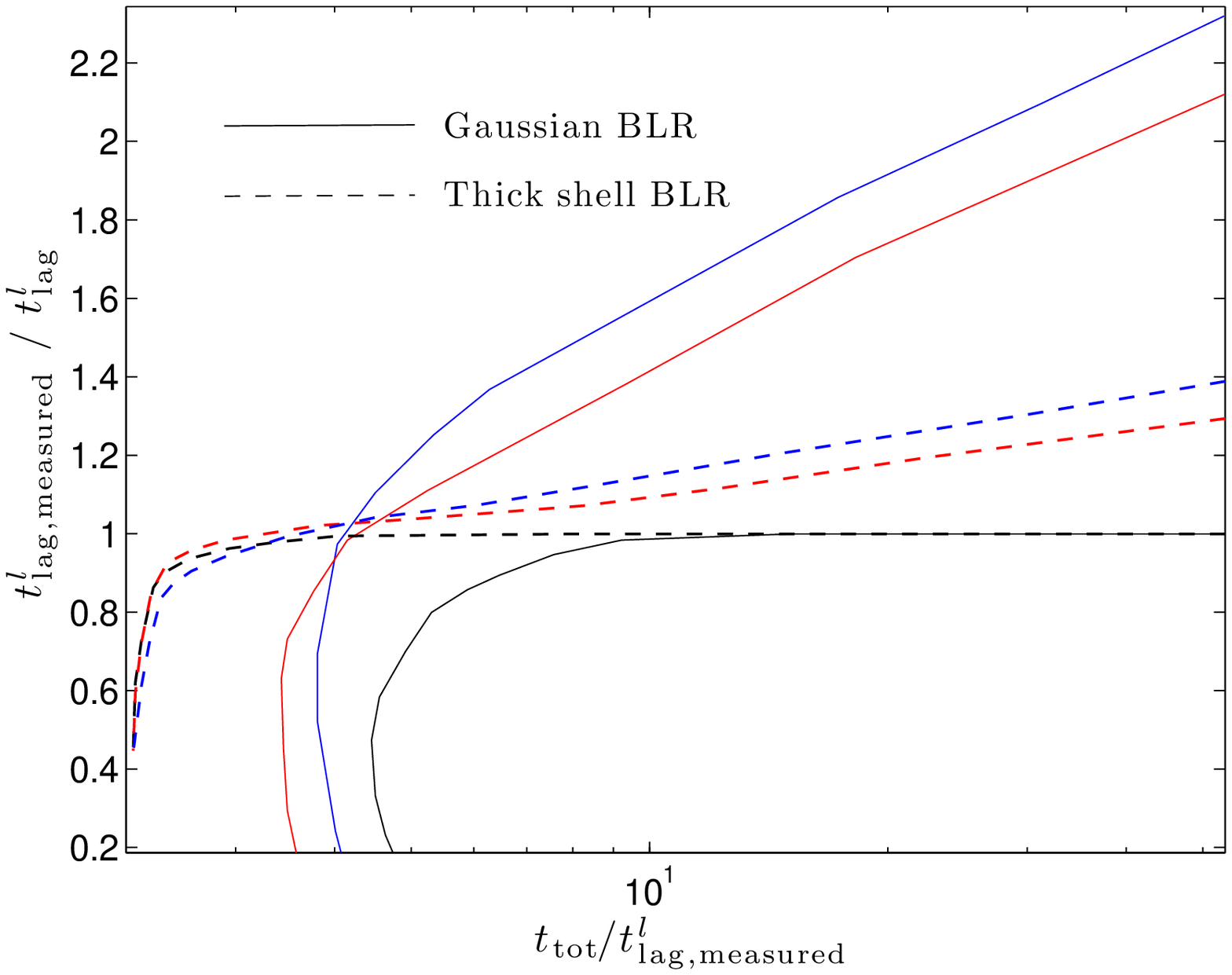}{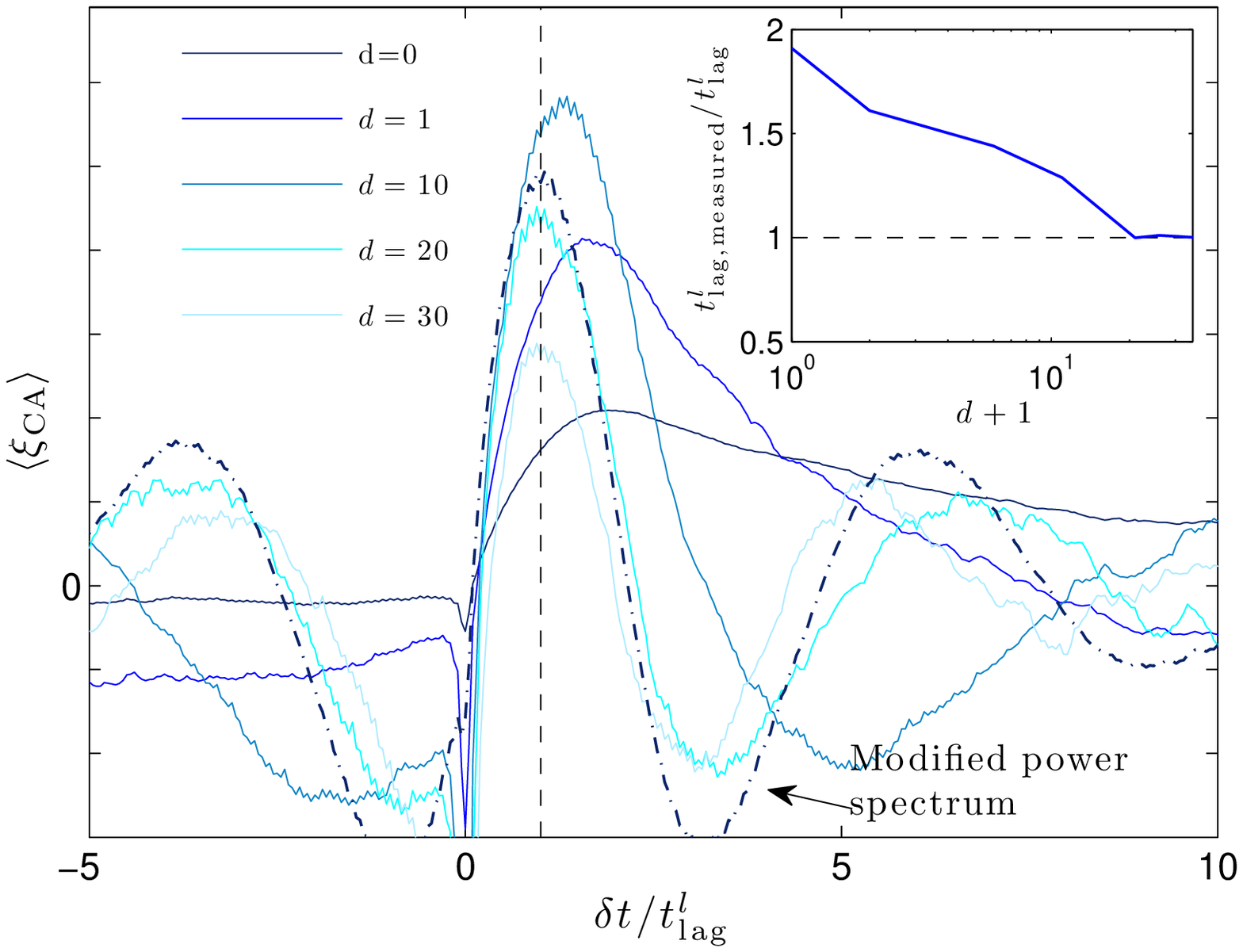}
\caption{{\it Left:} the bias in the time-lag measurement as a function of the observable quantity $t_{\rm tot}/t_{\rm lag,measured}^l$ ($\xi_{\rm CA}/\xi_{\rm AA}/\xi_{lc}$ results are shown in blue/red/black curves). For all statistical estimators there is a minimum $t_{\rm tot}/t_{\rm lag,measured}^l$ below which the sampling is inadequate. Above this threshold, the spectroscopic method converges to the input time lag (e.g., for a gaussian transfer function, $t_{\rm lag,measured}^l\simeq t_{\rm lag}^l$ for $t_{\rm tot}/t_{\rm lag,measured}^l>3$). In contrast, there is a bias using either photometric method wherein $t_{\rm lag,measured}^l/t_{\rm lag}^l$ is a rising function of $t_{\rm tot}/t_{\rm lag,measured}^l$, and is more pronounced for geometrically thick BLR configurations. {\it Right:} a particular example for $\left < \xi_{\rm CA}(\delta t) \right >$, calculated for an ensemble of 100 statistically identical objects with $t_{\rm tot}/t_{\rm lag}^l=50$, resulting in a peak at $\delta t \simeq 2 t_{\rm lag}^l$ (dark-black solid line). When the light curves of both bands are de-trended by a polynomial of degree $d$, the bias is gradually removed. For large enough $d$-values, substantial power on relevant  timescales is removed, thereby affecting the significance of the results (see legend). When the quasar power-spectrum is truncated at frequencies $\omega < \omega_{\rm min}$ [here $(2\pi/\omega_{\rm min})/t_{\rm tot}\simeq 0.13$],  the light curve appears to be stationary, and the bias disappears (dash-dotted dark blue line).}
\label{biases1}
\end{figure*}

Sampling is rarely uniform. In fact, seasonal gaps could be of the order of the time lag in some objects \citep{kas00}. The full treatment of sampling related issues is beyond the scope of this paper, and we restrict our discussion to light curve gaps (real data are treated in \S4). To check for potential systematics, we resampled our simulated light curves with equal "on" and "off" observing periods with durations $t_{\rm gap}$. We identify a bias when $t_{\rm gap}\simeq t_{\rm lag}^l$. Nevertheless, both the spectroscopic and the photometric methods are equally affected by it so that their results remain consistent. As our aim is to focus on cases in which the photometric approach leads to new or different biases, we do not further quantify this bias\footnote{See \S4 for the usage of discrete correlation functions, which are less susceptible to sampling issues.}.

\subsubsection{Inter-band continuum time delays}

Thus far, $t_{\rm lag}^c \ll t_{\rm lag}^l$ was assumed. While the spectroscopic method is unaffected by inter-band continuum delays (the continuum in the immediate vicinity of the emission line is considered), our definition of the continuum is that which corresponds to the filter having the least contribution of emission lines to its flux, i.e., the $X$-band. 

For a large set of simulation runs (obtained by varying both $\eta$ and $t_{\rm lag}^c/t_{\rm lag}^l$; not shown), we find that biases become non-negligible for $\eta^{-1}\vert t_{\rm lag}^c \vert /{t_{\rm lag}^l} \gtrsim 1$ ($t_{\rm lag}^c>0$ implies that continuum emission in the $Y$-band lags behind the $X$-band). Here we  consider particular cases in which $\vert t_{\rm lag}^c \vert/t_{\rm lag}^l \le 0.3$ ($\eta \simeq 0.13$ is assumed), and find that $\vert t_{\rm lag}^c\vert /t_{\rm lag}^l \gtrsim 0.1$ results in substantial, $>20\%$ bias (Fig. \ref{biases}) for the following reason: although modest values of $\vert t_{\rm lag}^c \vert /t_{\rm lag}^l$ are considered, the contribution of continuum processes to both light curves by far exceeds that of the emission line. Specifically, when the continuum in the $Y$-band lags behind the $X$-band, the deduced time lag reflects more on $t_{\rm lag}^c$ than on $t_{\rm lag}^l$.  When the $X$-band lags behind the $Y$-band, and $t_{\rm lag}^c/t_{\rm lag}^l$ is significant, the peak in both statistical estimators is strongly suppressed and time-lag measurements become less reliable. Realistically, inter-band continuum time-delays seem to be of order $10^{-2}t_{\rm lag}^l$ \citep{bac09}, and it is therefore likely that biases of the type discussed here would be negligible for most strong broad emission lines.

\subsubsection{Multiple emission lines}

Realistic quasar spectra could results in more than one emission line contributing to the $Y$-band, or even different emission lines contributing to both bands. While such complications are irrelevant for the spectroscopic method (note the small statistical scatter around unity in Fig. \ref{biases}), the photometric method is only sensitive to the combined contribution of all emission processes in some waveband. Here we study the biases introduced by including the contribution of a weaker secondary emission line to either of the bands whose time lag, $t_{\rm lag,s}^l \ne t_{\rm lag}^l$.

We find potentially considerable biases when the relative contribution of the weaker line to the flux, $\eta_s$ satisfies $\eta_s/\eta \lesssim 1$. However, the magnitude of the bias is also affected by $(t_{\rm lag,s}^l - t_{\rm lag}^l)/t_{\rm lag}^l$. We do not attempt to fully cover the parameter space in this work, and focus on particular cases. Figure \ref{biases} shows an example with $t_{\rm lag,s}^l/t_{\rm lag}^l=0.5$ where we find $\gtrsim20\%$ biases for $\eta_s/\eta \gtrsim 0.5$: when the weaker line contributes to the $Y$-band, $t_{\rm lag,measured}^l < t_{\rm lag}^l$ while the opposite occurs when the weaker line contributes to the $X$-band. Comparing to a case in which $t_{\rm lag,s}^l=t_{\rm lag}^l/10$ and $\eta_s/\eta=0.8$, we find similar trends but an overall smaller ( $<5\%$) bias. Assuming instead $t_{\rm lag,s}^l=2t_{\rm lag}^l$ and $\eta_s/\eta=0.8$, we find a bias of order 20\% (10\%) when the weaker line contributes to the $Y$- ($X$-) band. While the magnitude of the bias is clearly not solely determined by $\eta/\eta_s$, we find that the bias is, generally, small for $\eta_s/\eta \ll 1$. Clearly, if $t_{\rm lag,s}^l \ll t_{\rm sam}$ or $t_{\rm lag,s}^l \gg t_{\rm tot}$ then the time-series is insensitive to the presence of the secondary line, and the bias would be negligible regardless of $\eta_s/\eta$.

Further quantifying the bias as a function of the line properties (e.g., transfer functions and time-delays), filter response functions, quasar redshifts, and observing patterns is beyond the scope of this paper, and should be carried out on a case by case basis.

\subsubsection{Total light curve duration}

It is well known that the ratio of the total duration of the light curve to the time-lag, $t_{\rm tot}/t_{\rm lag}^l$, needs to be greater than some value to be able to adequately sample the transfer function and reliably measure the time lag using spectroscopic means. Our calculations demonstrate that the same holds also for the photometric method. In particular, for the gaussian (thick shell) transfer function, reliable measurements require that $t_{\rm tot}/t_{\rm lag}^l\gtrsim 3~(9)$; see figure \ref{biases1}.  

Differences between the photometric and the spectroscopic approach arise for $t_{\rm tot}/t_{\rm lag}^l \gg 1$: while the spectroscopic method gradually converges to $t_{\rm lag}^l$ (assuming sufficient data and adequate sampling; see Fig. \ref{biases1}), the photometrically deduced time lag, $t_{\rm lag,measured}^l$, develops a bias with respect to $t_{\rm lag}^l$ with increasing $t_{\rm tot}/t_{\rm lag}^l$. Generally, the bias is comparable (but not identical) for $\left < \xi_{\rm CA} \right >$ and $\left < \xi_{\rm AA} \right >$, and over-estimate the true lag. We quantify the bias for the two line transfer functions, $\psi^l$, defined above (\S2.1.2), and find it to be different in each case: broader $\psi^l$ that extend over larger $\delta t$ intervals, lead to more biased results. For example, while the thick shell BLR model shows a factor $\sim 2$ bias for $t_{\rm tot}/t_{\rm lag,measured}^l=50$, the corresponding bias for the Gaussian BLR is only $\sim 40$\% (see Fig. \ref{biases1}). The bias is seen to grow logarithmically with $t_{\rm tot}/t_{\rm lag,measured}^l$, and so would be most prominent when analyzing long time-series of low-luminosity quasars. Considering, for example, $\sim 10$\,years of LSST data for a low-$z$ quasar with a thick shell BLR whose size is $\sim 100$ light-days. In this case, $t_{\rm tot}/t_{\rm lag}^l\sim 30$, which would result in $t_{\rm lag,measured}^l/t_{\rm lag}^l\sim 2$. Not correcting for this bias would lead to an over-estimation of the black hole mass, by a similar factor, with considerable implications for black hole formation and evolution.

Consider the bias in  $\left < \xi_{\rm CA} \right> $: as its value is sensitive to the line transfer function, its origin may be traced to the $f_Y* f_X$ term in Eq. \ref{theta}, which is akin to the bias discussed in \citet{wel99}. In a nutshell, with the quasar's power spectrum being red, most of the power is associated with the lowest frequency modes. Nevertheless, those are the same modes that are not adequately sampled by any finite time series. As such, the data  {\it appear} to originate in a non-stationary process where, for example, the mean flux changes substantially between the extreme ends of the light curve. This and the fact that emission lines' contribution to the flux is always additive results in a positive bias. More extended line transfer functions, resulting in longer term correlations between the light curves, are therefore more prone to such biases. With  $\left < \xi_{\rm AA}(\delta t) \right> \simeq  \left < \xi_{\rm CA}(\delta t) \right> $, similar biases are also encountered there, that trace back to the $f_Y * f_Y$ term.

\begin{table*}
{
\begin{center}
\caption{Photometric properties of the PG sample of quasars$^{(a)}$}
\begin{tabular}{lllllllllllllll}
\tableline
Object		& 	& $\lambda L_\lambda(5100{\rm \AA})$ &    & B  &    &   &   R  & & net  & net  &  H$\alpha$ lag$^{(b)}$ & Predicted lag & Photo. lag\\
Name & $z$ & ($10^{44}\,{\rm erg~s^{-1}}$) & pts & $\chi_B$ & $\left < \delta_B \right>$ & pts & $\chi_{R}$ & $\left < \delta_R \right>$  & wo/Fe   & w/Fe & (days) & (days) & (days) \\
(1) & (2) & (3) & (4) & (5) & (6) & (7) & (8) & (9) & (10)  & (11)  & (12)  & (13) & (14) \\
\tableline
PG\,0026+129 & 0.142 & 	$7.0\pm1.0$		& 70		& 0.19	& 0.02	& 71	& 0.14	& 0.02 	& 0.03	& 0.07 	& $132^{+29}_{-31}$ & $147\pm25$ &-  \\
PG\,0052+251 & 0.155 & 	$6.5\pm 1.1$		& 75		& 0.26 	&0.03	& 75	& 0.20	& 0.02	& 0.04	& 0.08 	& $211^{+66}_{-44}$ & $141\pm27$ & - \\
{\bf PG\,0804+761} & {\bf 0.100} & $ {\bf 6.6\pm 1.2}$	& {\bf 83}		& {\bf 0.17}	& {\bf 0.01}	& {\bf 88}	& {\bf 0.14}	& {\bf 0.01}	& {\bf 0.05}	& {\bf 0.07} 	& ${\bf 193^{+20}_{-17}}$ & ${\bf 136\pm27}$ & ${\bf 200 \pm 100}$\\
{\bf PG\,0844+349} & {\bf 0.064} & 	${\bf 1.72\pm 0.17}$ 	& {\bf 65}		& {\bf 0.10}	& {\bf 0.02}	& {\bf 66}	&{\bf 0.10}	&{\bf 0.02}	&{\bf 0.08}	& {\bf 0.06}	& ${\bf 39^{+16}_{-16}}$ & ${\bf 51\pm6}$ & ${\bf 500\pm 200 }$ \\
PG\,0953+414 & 0.239 & 	$11.9\pm1.6$		& 60		& 0.14	& 0.03	& 60	& 0.12	&0.01	&0.06	& 0.11	& $187^{+27}_{-33}$$^{(c)}$ & $231\pm39$ & - \\
PG\,1211+143 & 0.085 & 	$4.93\pm0.80$	 	& 25		& 0.16	&0.01	& 24	&0.13	&0.01	& 0.06	& 0.07	& $116^{+38}_{-46}$ & $109\pm20$ & - \\
PG\,1226+414 & 0.158 & 	$64.4\pm7.7$		& 45		& 0.12 	&0.01	& 45	&0.10	&0.01	&0.04 	& 0.08	& $514^{+65}_{-64}$ & $703\pm135$ & - \\
{\bf PG\,1229+204} & {\bf 0.064} & 	${\bf 0.94\pm0.10}$		& {\bf 44}		& {\bf 0.16}	& {\bf 0.03}	& {\bf 45}	& {\bf 0.16}	& {\bf 0.03}	& {\bf 0.08}	& {\bf 0.06}	& ${\bf 71^{+39}_{-46}}$ & ${\bf 34\pm4}$ & {\bf N/A} \\
PG\,1307+085 & 0.155 & 	$5.27\pm0.52$		& 30		&0.14 	&0.01	& 30	&0.10	&0.01	&0.04	& 0.08	& $179^{+94}_{-145}$ & $122\pm16$ & - \\
PG\,1351+640 & 0.087 & 	$4.38\pm0.43$	 	& 35		&0.11 	&0.01	& 35	&0.10	&0.01	&0.06	& 0.07	& $247^{+162}_{-78}$ & $101\pm13$ & - \\
PG\,1411+442 & 0.089 & 	$3.25\pm0.28$	 	& 29		&0.10 	&0.01	& 29	&0.07	&0.01	&0.06	& 0.07	& $103^{+40}_{-37}$ & $82\pm9$ &- \\
PG\,1426+015 & 0.086 & 	$4.09\pm0.63$	 	& 26		&0.16	&0.01	& 25	&0.13	&0.02	&0.06	& 0.07	& $90^{+46}_{-52}$ & $96\pm17$ &- \\
{\bf PG\,1613+658} & {\bf 0.129} & 	${\bf 6.96\pm0.87}$	 	& {\bf 66}		&{\bf 0.11}	&{\bf 0.01}	& {\bf 64}	&{\bf 0.10}	&{\bf 0.02}	&{\bf 0.04}	& {\bf 0.07} & ${\bf 43^{+40}_{-22}} $ & ${\bf 144\pm22}$ &  ${\bf 300 \pm 130}$ \\
PG\,1617+175 & 0.114 & 	$2.37\pm0.41$	  	& 56		&0.19	&0.02	& 56	&0.14	&0.02	&0.04	& 0.07	& $111^{+31}_{-37}$ & $67\pm12$ &- \\
{\bf PG\,1700+518} & {\bf 0.292} & 	${\bf 27.1\pm1.9}$	 	& {\bf 53}		&{\bf 0.07}	&{\bf 0.02}	& {\bf 53}	&{\bf 0.07}	&{\bf 0.01}	&{\bf 0.05}	& {\bf 0.11} & ${\bf 114^{+246}_{-235}}$${\bf ^{(c)}}$ & ${\bf 428\pm61}$ & ${\bf 300 \pm 160}$\\
{\bf PG\,1704+608} & {\bf 0.371} & 	${\bf 35.6\pm5.2}$	  	& {\bf 38}		&{\bf 0.13}	&{\bf 0.01}	& {\bf 40}	&{\bf 0.11}	&{\bf 0.01}	&{\bf 0.02}	& {\bf 0.05}	& ${\bf 437^{+252}_{-391}}$${\bf ^{(c)}}$ & ${\bf 550\pm109}$ & ${\bf 400\pm200}$ \\
{\bf PG\,2130+099} & {\bf 0.061} & 	${\bf 2.16\pm0.2}$	 	& {\bf 78}		&{\bf 0.10}	&{\bf 0.02}	& {\bf 79}	&{\bf 0.08}	&{\bf 0.02}	&{\bf 0.08}	& {\bf 0.06}	&${\bf 237^{+53}_{-28} }$ & ${\bf 60\pm7}$ & ${\bf 240\pm100 }$ \\
\tableline
\end{tabular}
\label{tab}
\end{center} }
{ Columns: (1) Object name, (2) redshift, (3) monochromatic luminosity in units of $10^{44}\,{\rm erg~s^{-1}}$, (4)-(6) properties of the photometric light curve of the $B$-band: number of points, the normalized variability measure, and the mean measurement uncertainty. (7)-(9) having the same meaning as (4)-(6) but for the $R$ band. (10) the net contribution of emission lines to the bands excluding the iron blends. (11) like (10) but including the iron emission complexes. (12) the spectroscopically measured time lag for the H$\alpha$ line from \citet{kas00}. (13) The {\it observed} lag for the Balmer emission lines as predicted by the BLR size-luminosity relation of \citet[see their Eq. 6]{kas00}. (14) The time lag measured in this work using broadband light curves.\\
(a) Highlighted table rows indicate objects for which a statistically significant  signal is detected (see \S4.3).\\
(b) Values correspond to centroid observed-frame values \citep[see their table 6]{kas00}. \\
(c) Data for H$\alpha$ are not available. Time lag corresponds to the H$\beta$ line.}
\end{table*}

To verify the source of the bias, we repeated the calculations using a truncated form for the power-spectrum where all frequencies below a minimum frequency, $\omega_{\rm min}$, have no power associated with them (we have experimented with $2\pi/t_{\rm lag} \gg \omega_{\rm min} \gg 2\pi/t_{\rm tot}$). The resulting $\left < \xi_{\rm CA}(\delta t) \right> $ is indeed unbiased and peaks at $t_{\rm lag}^l$ (not shown). For intermediate cases where, for example, the power spectrum flattens at low frequencies \citep{cze99}, the results will be less biased but may not be entirely bias-free.

Having identified the source of the bias, it is now possible to correct for it using one of several methods which we term:  de-trending, sub-sampling, and bootstrapping. De-trending has been shown to be useful to correct for biases in the spectroscopic method \citep{wel99}. Basically, one filters out long-term trends in the light curve by fitting (done here in the least-squares' sense) and subtracting off a polynomial of some degree, $d$, from the data. As a polynomial of degree $d$ has $d$ zeros, subtracting a higher degree polynomial from the data results in a suppressed variance at frequencies $\omega \lesssim d (2\pi/t_{\rm tot})$. Conversely, the highest degree polynomial that can be subtracted from the data, whilst not suppressing the reverberation signal, is $d\lesssim \lfloor t_{\rm tot}/t_{\rm lag} \rfloor $. The effect of de-trending is shown in figure \ref{biases1} for a range of $d$-values, and assuming $t_{\rm tot}/t_{\rm lag}^l=50$. While subtracting a $d=2$ polynomial already decreases the bias substantially, a bias-free result is recovered by de-trending with a $d=20~(=0.4 t_{\rm tot}/t_{\rm lag}^l)$ polynomial. Further increasing $d$ suppresses the sought after signal to insignificant levels by removing modes with frequencies $\sim 2\pi/t_{\rm lag}^l$. The dependence of the remaining bias on the de-trending polynomial degree, $d$, is shown in the inset of figure \ref{biases1}. 

An alternative method to remove the bias is to analyze continuous sub-samples of the time series, thereby reducing the effective $t_{\rm tot}$ and decreasing the bias. This method can be implemented in cases where one is not data-limited. 

The third method for removing the bias involves bootstrapping: given an observed continuum light curve (i.e., $f_X$), and an assumed $\psi^l$, the light curve of the line-rich band is simulated. Upon sampling it using the cadence of the real data, an artificial $f_Y$ light curve is obtained. By calculating the statistical estimators for the semi-artificial data, it is possible to compare $t_{\rm lag,measured}^l$ to the input lag, and correct for the bias. Conversely, should the bias be known by other means, inferences concerning the nature of the line transfer function may be obtained.

\section{Data Analysis}

Having gained some understanding of the advantages and limitations of broadband photometric reverberation mapping, we now wish to test the method using  real data. Unlike the models described in \S3, real data are often characterized by uneven sampling, gaps, and a varying signal to noise across the light curve. Here we consider a subset of the Palomar-Green (PG) sample of objects\footnote{We choose to not focus our attention on active galactic nuclei (i.e., low luminosity quasars) since, in this case, great care is needed to make sure that varying seeing conditions do not introduce additional noise whose characteristics are poorly understood. In this case, specific data reduction techniques, such as aperture photometry, may be of advantage.} \citep{sg83}, for which Johnson-Cousins $B$ and $R$ photometry is publicly available\footnote{http://wise-obs.tau.ac.il/PG.html} \citep{mao94,giv99}, and  independent BLR size measurements are available through spectroscopic reverberation mapping \citep{kas00}. The properties of the sample are given in table 1. 

While the PG sample provides a natural testbed for our purpose, it is far from being ideal: the typical number of points per light curve is small, $\sim 50$, and the photometric errors non-negligible, $\lesssim2$\% (c.f. Fig. \ref{1sigma}). Other photometric data sets are, in principle, available  \citep{geh03,meu11} but do not yet have BLR size measurements with which our findings may be compared [see  \citet{che11}]. 

\subsection{Preliminaries}

\begin{figure*}
\plottwo{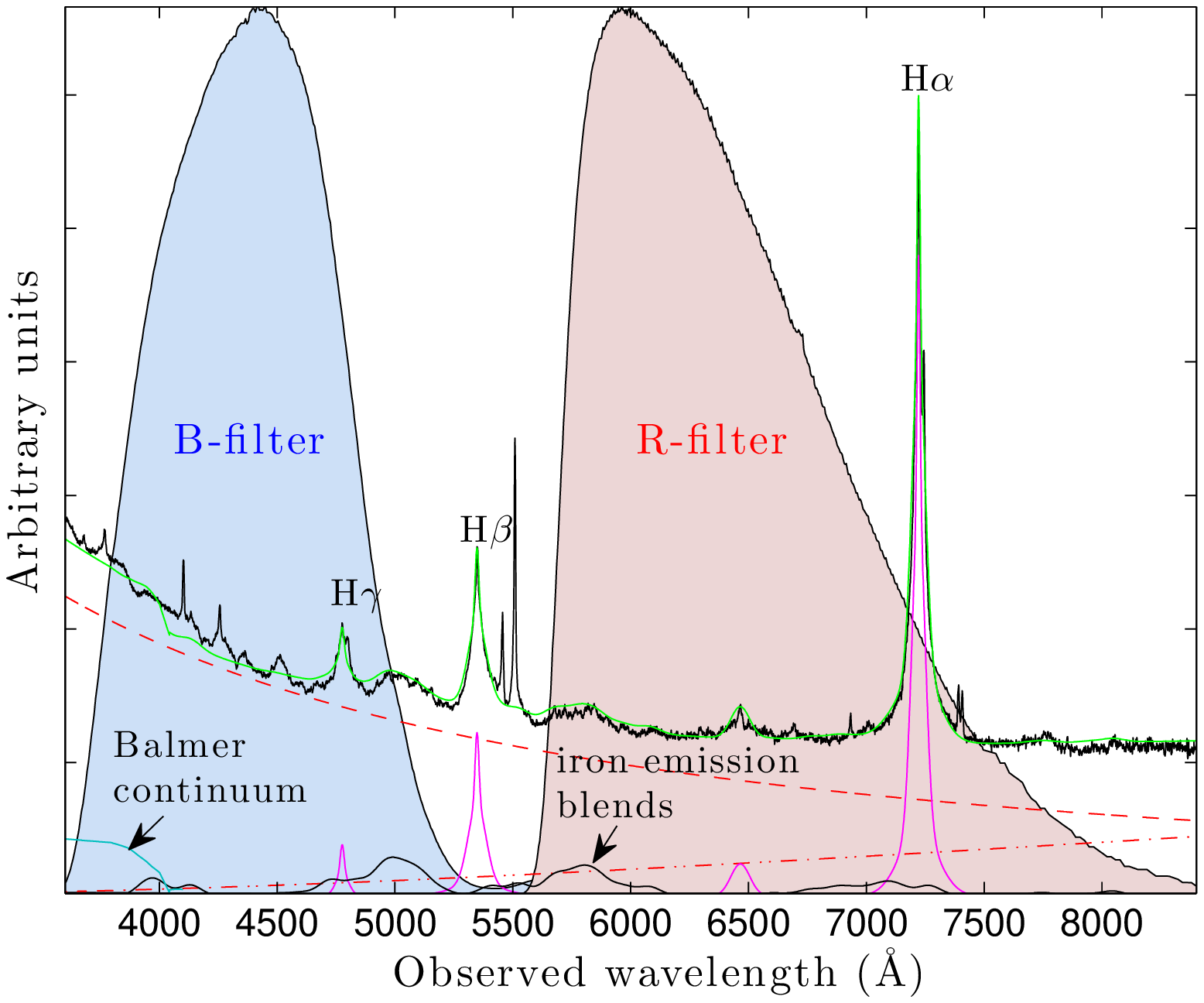}{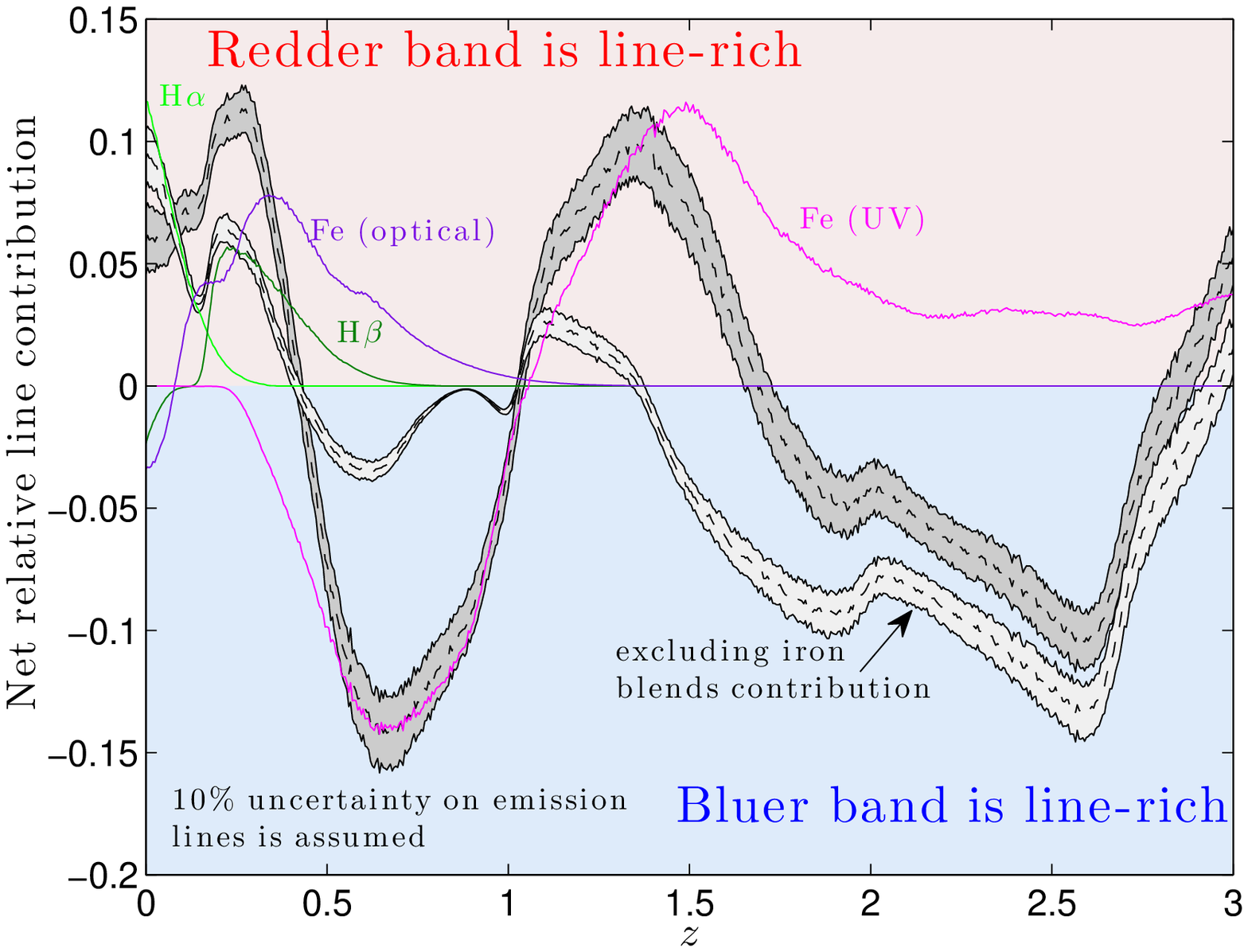}
\caption{Emission line and continuum contribution to the Johnson-Cousins broadband filters. {\it Left:} the composite quasar spectrum \citep[assumed to be at $z=0.1$; see table \ref{tab}]{dvb01} is shown together with the $B$- and $R$-filter response functions. Spectral decomposition into prominent emission lines (magenta) and iron line blends (gray curve) is shown. A double powerlaw continuum emission model is also included (dashed and dash-dotted magenta lines), and the "best-fit" model is shown in green. {\it Right:} the net emission line contribution to the broadband light curves as a function of the quasar redshift (dashed black lines; gray hatched patterns indicate the uncertainties assuming a 10\% uncertainty on the flux of individual lines). The net contribution of a few particular lines is shown in colored curves. The sum of all lines excluding iron (light grey) and including iron (dark grey) are shown, indicating that in either case the $R$-filter is relatively line-rich for $z<0.1$ objects with the Balmer emission line contribution being the dominant one.  Typical redshift intervals over which line-rich filter identifications change are greater than the typical uncertainty associated with photometric redshifts \citep{wei04,wu10}.}
\label{wise}
\end{figure*}

We first identify line-rich and line-poor bands given the typical quasar spectrum \citep{dvb01},  the spectroscopically-determined redshift for the PG sample \citep{kas00}, and the relevant filter throughput curves (Fig. \ref{wise}). In particular, we need to identify the filter with the larger relative contribution of emission lines to the variance of the light curve. This is, however, difficult to estimate, since the variability properties of only a few lines, in only a few objects, are sufficiently well understood. Instead, we adopt a simplified approach and {\it assume} that the contribution of emission lines to the variance is proportional to their relative  contribution to the flux. This is a fair approximation if, for example, the relative flux variations in all emission lines are comparable. This is supported by theory \citep{kas99} and by the fact that the fractional variability of the Balmer lines is independent of their flux \citep[see their table 5]{kas00}. 

To properly account for the contribution of emission lines to the flux in some band, spectral decomposition of all prominent line, line blends, and continuum emission processes is required. To this end, we considered the high signal-to-noise (S/N) composite quasar spectrum of \citet{dvb01}, and identified all prominent emission lines and blends in the relevant wavebands. To fit for continuum emission, we identified line-free regions over the entire spectral range from rest-UV to the optical \citep{kas00,dvb01}, and fitted a double powerlaw model (see Fig. \ref{wise}), which accounts for the changing slope at optical wavelengths. The stronger Balmer emission lines were fitted with a two Gaussian model, which accounts for their relatively narrow line cores and extended wings. Weaker lines were fitted with a single Gaussian model. Iron blend emission templates, as derived for IZw I \citep{ver04,ves01}, were convolved with a Gaussian kernel and fitted to the data. Independently adjusting the width and normalization for each Gaussian kernel (individual lines and blends), a reasonable agreement was sought. We model the Balmer continuum using the analytic expression of \citet{gra82}. We emphasize that our model is by no means physical, nor uniquely-determined, and that its sole purpose is to allow for the qualitative estimate of the emission line contribution to the flux in each band. The decomposed spectrum is shown in figure \ref{wise}.

The line-rich band, as a function of the quasar redshift, for the Johnson-Cousins filter scheme, is shown in figure \ref{wise}. Clearly, depending on the object's redshift, either the $B$ or the $R$ filter may be identified with the line-rich band. Typically, a redshift accuracy of 0.2 is required to securely identify the line-rich band, which is within reach of  photometric redshift determination techniques \citep{ric01}. We find that, for $z<0.5$ quasars, the $R$ band may be identified with the line-rich band (i.e., with the $Y$-band, using the nomenclature of \S2.1.2). This result is rather insensitive to whether or not the optical iron blends contribute to the variance on the relevant timescales. For $z<0.25$ objects, the flux and, by assumption, the variance, are largely dominated by the Balmer lines. Uncertainties of order 10\% in the flux of individual lines and line blends do not significantly change this result. 

\subsection{Methods}

Aiming to analyze individual objects in the PG sample having a marginally adequate number of points in their light curves (thereby resulting in a large uncertainty on the determined lag; Fig. \ref{1sigma}), we outline the formalism used below to assess the significance of our results.

\subsubsection{Cross-correlation analysis}

\begin{figure*}
\epsscale{1.2}
\plotone{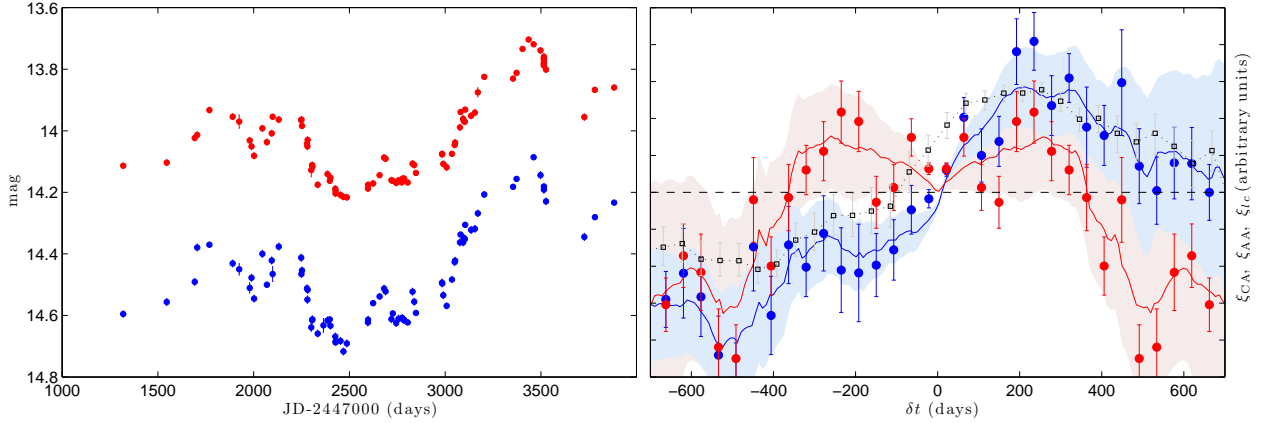}
\caption{Light curve analysis for PG\,0804+761. {\it Left:} The $B$ and $R$ light curves show significant non-monotonic variability over a few hundreds days timescales (see table 1). {\it Right:} the calculated statistical estimators, $\xi_{\rm CA}$ (blue shades) and $\xi_{\rm AA}$ (red shades). Also shown is the renormalized $\xi_{lc}$ (black points) based on the spectroscopic data of \citet{kas00}. Two analysis methods are used to calculate the photometric estimators: ICCF (solid lines and shades) and ZDCF (discrete points), which lead to consistent results. The uncertainty on each estimator was calculated using 100 Monte Carlo simulations: the FR-RSS method was used to estimate the error on the ICCF, and a random independent pair selection was used to estimate the error on the ZDCF (see text). All estimators peak at $\delta t \simeq 200$\,days. The result obtained using $\xi_{\rm CA}$ has a higher statistical significance, and both photometric estimators are less significant than the spectroscopic one.}
\label{0804}
\end{figure*}

In analyzing real data for individual objects in the PG sample (table 1), it is important to verify that the detected signal is real and is not some artifact of non-even sampling or the particular light curve observed. There are several methods in the literature for cross-correlating realistic data sets: the interpolated cross-correlation function \citep[ICCF; used in \S3]{gas86} and the discrete correlation function \citep[DCF]{edl88}. While the ICCF may be somewhat more sensitive to low-level signals, it is also more prone to sampling artifacts \citep{pet93}. To verify the significance of the ICCF signal, two algorithms are often used to estimate the uncertainty in the deduced correlation function: the flux randomization (FR) and the random subset selection (RSS). In a nutshell, the first algorithm adds noise according to the measurement errors, and repeats the analysis many times to estimate the uncertainty associated with the results, while the latter method resamples the light curves (while losing $\sim 37$\% of the data), to reach the same goal.  Quite often both tests are combined to form the FR-RSS method, which is our method of choice in the present analysis. When considering the DCF approach, the $z$-transformed DCF algorithm (ZDCF) by \citet{ale97} often produces the best results, and is the approach adopted here. While all the aforementioned algorithms have been thoroughly tested with respect to spectroscopic reverberation mapping, their adequacy for the photometric reverberation mapping method is yet to be examined.

Our implementation of the ICCF (FR and RSS algorithms alike) is standard \citep{pet98}. Nevertheless, because photometric reverberation mapping requires the subtraction of two cross-correlation functions, and to avoid interpolations of any kind\footnote{Although we do not use this method here, we note that due to the correlated nature of the correlation function, interpolation at the correlation function level is likely to provide results which are less sensitive to sampling artifacts than the ICCF method.}, our implementation of the ZDCF algorithm requires that both the $f_Y * f_X$ and $f_X * f_X$ terms in equation \ref{theta} are evaluated over an identical time grid, and that the bin center is identified with the time tag for that bin, i.e. $\delta t$. Similar considerations apply also when evaluating $\xi_{\rm AA}$. Furthermore, we require that the number of independent pairs contributing to each bin in the correlation functions is $\ge 11$ \citep{ale97}. For simplicity, we use equally spaced bins to evaluate $\xi_{\rm CA}$ and $\xi_{\rm AA}$, and repeat the analysis many times by randomly selecting independent pairs \citep{ale97}. This provides us with a distribution of $\xi_{\rm CA}$ and $\xi_{\rm AA}$ values at each time bin, from which the mean and its uncertainty (associated with one standard deviation) are evaluated.

Spectroscopic reverberation mapping often employs two distinct approaches to estimate the time lag: by measuring the position of the peak of the correlation function, or by measuring its centroid. While the results are comparable, they are not identical and many works have discussed the pros and cons associated with each approach \citep[and references therein]{kas00}. Similar algorithms are implemented here. 

Estimating the uncertainty on the measured time-lag is done in the following way: we use the set of Monte Carlo simulations described above and calculate $\xi_{\rm CA},~\xi_{\rm AA}$ for each realization. The time-lag is identified either with the peak or with the centroid of the statistical estimators\footnote{Here we use the smoothing algorithm defined in \S3.1, to avoid identifying spurious peaks with the physically meaningful peak.}, and a distribution of time-lags is obtained. The uncertainty on the time-lag is associated with one standard deviation of that distribution. 

\subsubsection{Controlling systematics}

In addition to the above algorithms, aimed at testing the robustness of the signal, we provide an additional test to identify systematic effects of the type discussed in \S3.2.5 and to assist in screening against problematic data sets that  may lead to spurious signals. To this end we use the bootstrapping method discussed in \S3.2.5, verifying that the mean contribution of the simulated emission line to the flux of the line-rich band is that given in Fig. \ref{wise}, and that the variance of the line light curve is consistent with the \citet[see their table 5]{kas00} results.

\subsection{Results}

Here we outline the results for individual objects enlisted in table 1: we first discuss quasars for which a significant signal is detected\footnote{The significance criterion used here requires a $\gtrsim 1\sigma$ signal (with $\sigma$ being the one standard deviation uncertainty on the measured value) detected in both statistical estimators, and for which the ICCF and ZDCF analysis yield consistent results.}, and then cases where the analysis does not result in a discernible peak in either estimator. Our treatment of the ensemble as a whole is given in \S4.3.9.

\subsubsection{PG\,0804+761}

The $B$ and $R$ light curves for PG\,0804+761 (henceforth termed PG\,0804), at $z=0.1$, are shown in figure \ref{0804} where large non-monotonic flux variations are evident. This object has the best sampled light curves in the \citet{kas00} sample, with a mean uncertainty on the flux measurement in both bands being of order 1\% (with maximum uncertainties on individual measurements being $\sim 2$\%). The normalized variability measure is  among the largest in the sample $\sim 17\%/14\%$ for the $B/R$ bands (Table 1). As such, it is a promising candidate for photometric reverberation mapping.

\begin{figure}
\plotone{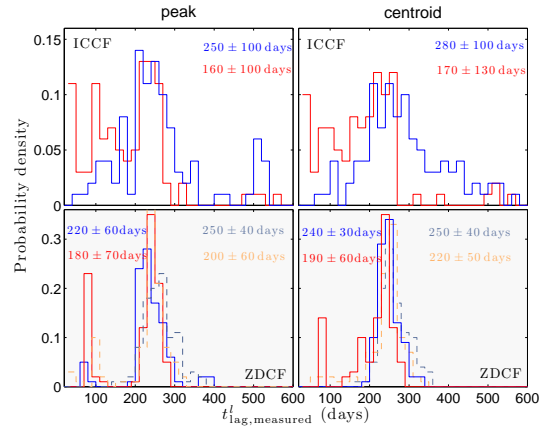}
\caption{The peak (left panels) and centroid (right panels) $t_{\rm lag,measured}^l$ distributions for PG\,0804,  obtained using sets of 100 Monte Carlo simulations (standard color scheme is used). ICCF (ZDCF) results are shown in the upper (lower) panels. When using the ZDCF approach, two different bin sizes were assumed (solid and dashed lines; red/blue shades correspond to $\xi_{\rm AA}/\xi_{\rm CA}$), and consistent results are obtained in both cases. The mean lag and its uncertainty (assumed to be one standard deviation) corresponding to each distribution are denoted in each panel.  Centroid statistics results in slightly larger time-lags due to the non-symmetric shape of the statistical estimators around the peak (Fig. \ref{acfccf}). While the time-lags inferred by all statistical estimators and methods are statistically consistent, the uncertainties inferred by the ZDCF approach are relatively small, and possibly less reliable (see text).}
\label{0804_dist}
\end{figure}

\begin{figure}
\plotone{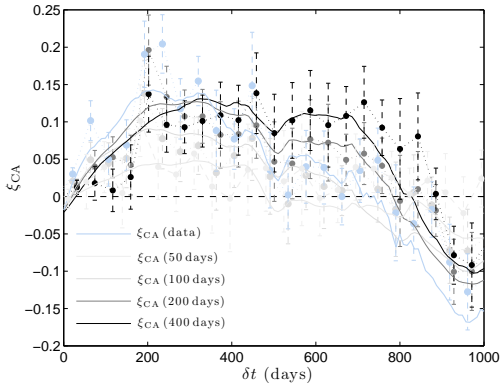}
\caption{Photometric analysis of real $B$ data and simulated $R$ data for PG\,0804. Several artificial time-lags (see legend) are assumed and the line-rich $R$ band light curve is simulated using the line-poor $B$-band light curve (see \S4.2.2). Analysis results for the real data are shown, for comparison, in blue shades. Both ICCF and ZDCF results are shown. Different artificial time-lags result in statistically different signals. As expected, longer time-lags result in a hump extending to longer timescales. The current data set is insensitive to short lags due to inadequate cadence. Qualitatively, an input time lag of 100-200\,days is consistent with the observed signal.}
\label{0804_sys}
\end{figure}

Figure \ref{wise} indicates that, at its redshift, the $R$ band is relatively line-rich  (the net contribution of the lines to the bands is $\sim 6$\%), with H$\alpha$ being the dominant emission line. We calculated $\xi_{\rm CA}$ and $\xi_{\rm AA}$ using the ICCF and ZDCF algorithms. Unsurprisingly, the ZDCF method is somewhat more noisy than the ICCF methods \citep[see their Fig. 4]{kas00}. That said, we find the results of all methods of analyses to be consistent given the uncertainties. This means that interpolation is not a major source of uncertainty for this object, which is consistent with the relative regular sampling of its light curves. 

\begin{figure*}
\epsscale{1.2}
\plotone{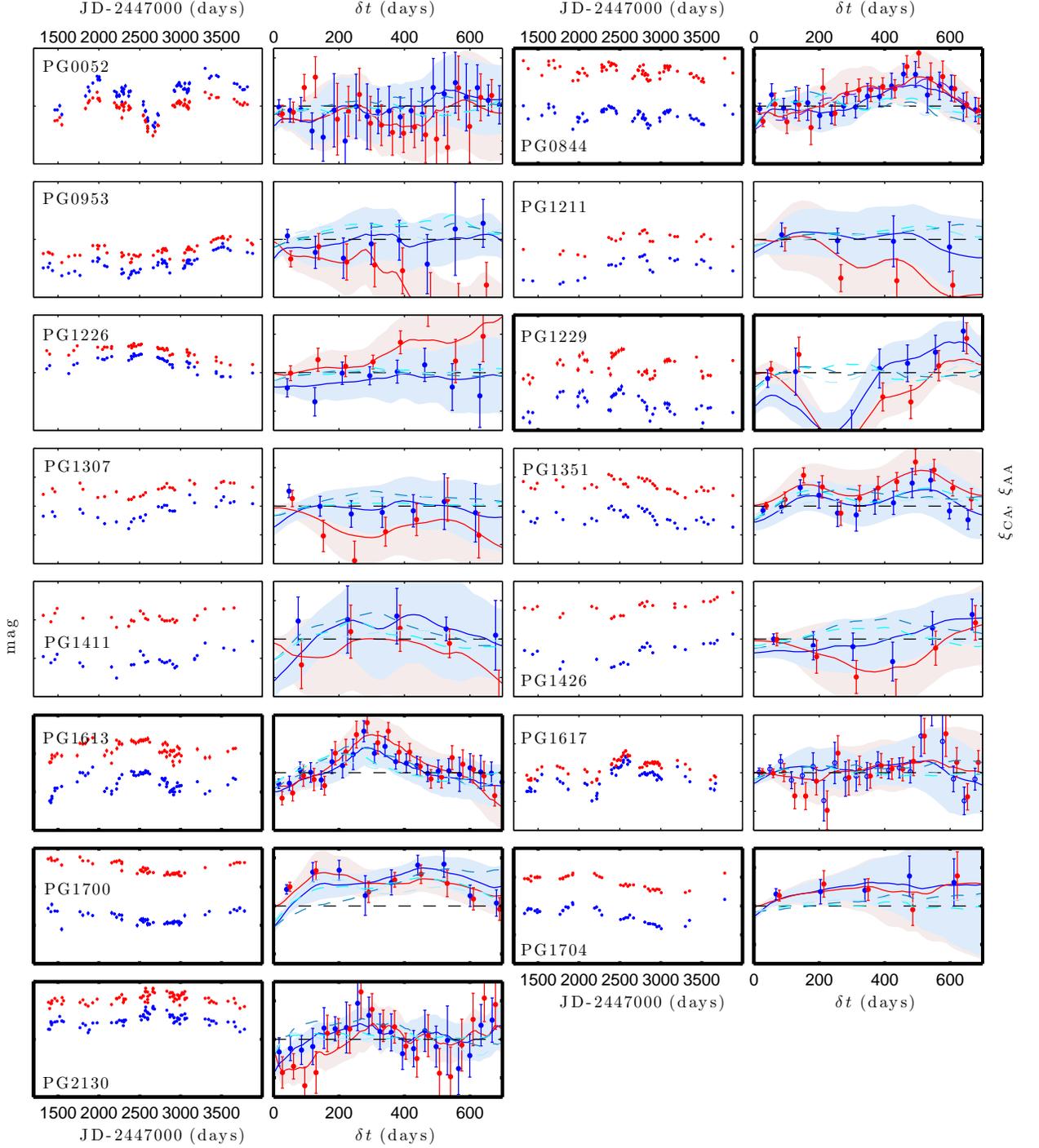}
\caption{Analysis of 15 PG quasars from the sample of \citet[see our table 1]{kas00}. In each pair of columns, the left panel shows the $B$ (blue points) and $R$ (red points) light curves, and the right panel the calculated $\xi_{\rm CA}$ (blue curves and shades) and $\xi_{\rm AA}$ (red curves and shades) using the ICCF (solid lines) and ZDCF (points with error-bars) numerical schemes. Cases for which a statistically significant signal is detected in both estimators and using both numerical schemes, are highlighted. For each quasar, the results of three $\xi_{\rm CA}$ simulations are shown (in dashed blue-shaded curves) based on the expected time-lag from the BLR size vs. luminosity relation, $t_{\rm lag, ex}^l$ (table 1): light to dark shades correspond simulated time-lags of $t_{\rm lag, ex}^l/2 \to t_{\rm lag, ex}^l \to 2t_{\rm lag, ex}^l$ (cyan curves correspond to $t_{\rm lag, ex}^l$). For PG\,0844, a simulation with a 500\,days lag is also included (dashed magenta line). The ICCF results reported here are used to construct the ensemble averages considered in figure \ref{pg_mean}.}
\label{all_pgs}
\end{figure*}

Our analysis shows the expected features: a hump in both $\xi_{\rm CA}$ and $\xi_{\rm AA}$, which peaks at similar ($\delta t \sim 200$\,days) timescales. For comparison, we used a published light curve for the broad H$\alpha$ emission line \citep{kas00}\footnote{see http://wise-obs.tau.ac.il/~shai/PG/}, and calculated $\xi_{lc}$ using the ZDCF algorithm. Clearly, both statistical estimators defined here trace the line-to-continuum cross-correlation function with a varying degree of significance: $\xi_{\rm CA}$ shows a $\gtrsim 2\sigma$ peak ($\sigma$ being the uncertainty around the maximum), while the peak in $\xi_{\rm AA}$ is only $\gtrsim 1\sigma$ significant.

Using the Monte Carlo simulations described in \S4.2.1, peak or centroid time distributions were obtained, and are shown in figure \ref{0804_dist}. Clearly, time-lag estimates qualitatively agree between the different methods. Quantitatively, however, there are some differences: the centroid method results in somewhat larger values for the time-lag. This results from the non-symmetric nature of $\xi_{\rm CA},~\xi_{\rm AA}$ showing a more extended wing toward longer time-delays (Fig. \ref{0804}). Results based on the ZDCF approach result in errors, which are typically smaller than those obtained using the ICCF method. This results from the fact that the ZDCF peak is largely determined by a few individual bins (note the two bins around 200\,days that lie above most other ZDCF points, as well as above the ICCF curve; Fig. \ref{0804}). In fact, the resulting uncertainty on the ZDCF, being of order 20\% in this case, is smaller than predicted based on our Monte Carlo simulations presented in figure \ref{1sigma}.  We conclude that estimating the uncertainty on the time-lag using the ZDCF results may be less reliable.

Using the ICCF time-lag distributions, we determine the observed time-lag for PG\,0804 to be $250\pm100$\,days, which is statistically consistent with the spectroscopically measured value of $193^{+20}_{-17}$\,days \citep{kas00}. As far as systematics is concerned, we do not expect a bias in the measured lag to exceed $\sim 40$\%  given that $t_{\rm tot}/t_{\rm lag,measured}^l\gtrsim 10$ (Fig. \ref{biases1}). 

Using the numerical bootstrapping scheme defined in \S4.2.2, we find that an input time lag of $100-200$\,days is qualitatively consistent with the data. In particular, much shorter or much longer time-lags, are inconsistent with our findings in terms of signal amplitude and shape. Interestingly, a weak yet marginally significant hump at around $200-300$\,days is implied by the data even when much shorter artificial time-lags are used as input. Therefore, one should be careful when interpreting marginally significant signals. For the specific case considered here we can conclude that the data cannot be used to detect short time-lags, which is of no surprise given that the cadence of the light curves is at $20-30$\,days intervals.

To conclude, the time-lag estimated here by the photometric reverberation mapping approach is $\simeq 200\pm100$\,days, and is therefore consistent with the value deduced by \citet{kas00}, albeit with an expectedly larger uncertainty.

\subsubsection{PG\,0844+349}

This low-$z$, low-luminosity object has a potentially prominent ($\sim 8\%$) line contribution to the $R$-band from the Balmer emission lines. With $\lesssim 70$ points in its light curve, and what appears to be significant variations over 200\,days time scales, our analysis reveals a marginal ($\gtrsim 1\sigma$ using the FR-RSS method and $\gtrsim 2$ using ZDCF statistics) peak at $\sim 500$\,days (Fig. \ref{all_pgs}). The spectroscopically measured time-lag for this object is of order 50\,days and agrees with the value expected from the BLR size vs. luminosity relation (table 1). Nevertheless, the data do not show a corresponding peak at those time-scales. In fact, our simulations indicate that only a time-lag $\gtrsim 100$\,days, could have been detected with some certainty. Interestingly, simulations with an input lag of 500\,days (\S4.2.2) are able to qualitatively account for the observed signal (dashed magenta curve in figure \ref{all_pgs}).  

It is not clear whether a 500\,days lag, {\it if} real, is associated with a long-term reverberation signal of the BLR, since there is no corresponding signal in the \citet{kas00} analysis of the spectroscopic data. We note, however,  that it is also possible that the measurement uncertainties are somewhat under-estimated for this object, hence the significance of our deduced signal over-estimated. That this might be the case is hinted by the normalized variability measure in both the $B$ and $R$ bands being similar (contrary to naive theoretical expectations) and the fact that contamination by the host galaxy in this low-$z$ quasar may be important (see below). 

\subsubsection{PG\,1229+204}

This quasar, being at low redshift ($z=0.064$) has one of the most prominent ($\lesssim 8\%$) Balmer line contributions to the $R$-band among the PG objects in our sample. The analysis detects a significant {\it trough} at $\gtrsim 200$\,days, in both statistical estimators, and using both the ICCF and the ZDCF methods. This is unexpected since a peak rather than a trough is predicted by the simulations for an expected lag of $\sim 34$\,days (Fig. \ref{all_pgs}). Simulating cases with time-lags of order 200\,days also produces a peak, in contrast to the observed signal. The interpretation that the $B$- rather than the $R$-band is the line-rich band, is not supported by the optical spectrum showing rather typical quasar emission lines \citep{kas00}. 

We cannot positively identify the origin of the unexpected signal but note that, at its low redshift, and being a rather faint low-luminosity quasar, the host galaxy is likely to substantially contribute to the flux in the bands. This has a dual result: 1) the actual contribution of the emission lines to the bands is smaller than that predicted here, and  2) the data may be substantially affected by varying seeing conditions between the nights, thereby contributing to the variance in the light curves, and masking the true signal. Interestingly, the reduced variability measure in both bands is similar for this object, which may indicate the presence of an additional source of noise not accounted for by the reported measurement uncertainties.

\subsubsection{PG\,1613+658}

Analyzing the data for this object, a marginally significant signal (at the $1\sigma-2\sigma$ level) is detected in both statistical estimators, at $\lesssim 300$\,days (Fig. \ref{all_pgs}). Formally,  $\xi_{\rm CA}$ statistics implies that the time lag is $300\pm130$\,days. At its redshift, the net contribution of emission lines is of order 5\%, and is therefore larger than the typical flux measurement uncertainty in this object ($\lesssim 2\%$). Simulations with an artificial lag of $\sim 300$\,days are able to qualitatively account for the observed signal. Much shorter time lags [e.g., $\sim 40$\,days, as spectroscopically deduced by \citet{kas00}] do not seem to be supported by our findings. Interestingly, for this object, spectroscopic time-lag measurements using the peak distribution function give very large errors for the H$\alpha$ line, and given its luminosity, PG\,1613+658 lies considerably below the BLR size vs. luminosity relation \citep[see also our table 1]{kas00}. Using our deduced lag, a better agreement is obtained with the \citet{kas00} size-luminosity relation.

\subsubsection{PG\,1700+518}

This object shows a statistically significant ($2\sigma-3\sigma$), wide peak extending over the range 150-450\,days. Assuming that H$\beta$ is the main line contributor to the $R$ band's flux, our simulations indicate that the signal is consistent with a time-lag of $\sim 430$\,days (formally, $\xi_{\rm CA}$ statistics yields a $300\pm160$\,days' lag). This value is statistically consistent with the time-delay measurement of \citet{kas00}, but our deduced lag is in somewhat better agreement with the expected BLR size given the source luminosity (table 1). Our simulations show that time-lags considerably shorter ($\ll 200$\,days) or longer ($\gg 800$\,days) are less favored by the data and cannot fully account for the observed signal (Fig. \ref{all_pgs}).

\subsubsection{PG\,1704+608}

This object, having $\sim 40$ visits in each band, shows a marginal ($\gtrsim 1\sigma$) hump at 200-300\,days. Formally, the deduced time-lag is $400\pm200$\,days, and is consistent with the values of \citet{kas00}.  At $z=0.371$, the net contribution of emission lines to the line-rich $R$-band is only about twice as large as the mean measurement uncertainty. This, and the small number of data points, account for the low-significance of the observed signal. Further, simulations show that a marginal signal is indeed expected to occur at roughly the spectroscopically determined lag.

\subsubsection{PG\,2130+099}

There is a hint for a localized ($\sim 1\sigma$) peak around 200-300\,days in both statistical estimators ($\xi_{\rm CA}$ statistics yields a lag of $240\pm100$\,days) using both the ICCF and the ZDCF methods. At face value, this timescale is consistent with the results of \citet{kas00} who find a time lag of $\sim 200$\,days. Nevertheless, as discussed by \citet{gr08}, the cadence of the \citet{kas00} observations may not be adequate to uncover the true time lag in this object. Specifically, \citet{gr08} find a time lag of order 30\,days using better sampled light curves on short time scales. The fact that we do not find a significant peak on short timescales is not surprising since we are essentially using the \citet{kas00} data set.

\subsubsection{PG\,0026+129 and other objects yielding null results}

\begin{figure*}
\epsscale{1.2}
\plotone{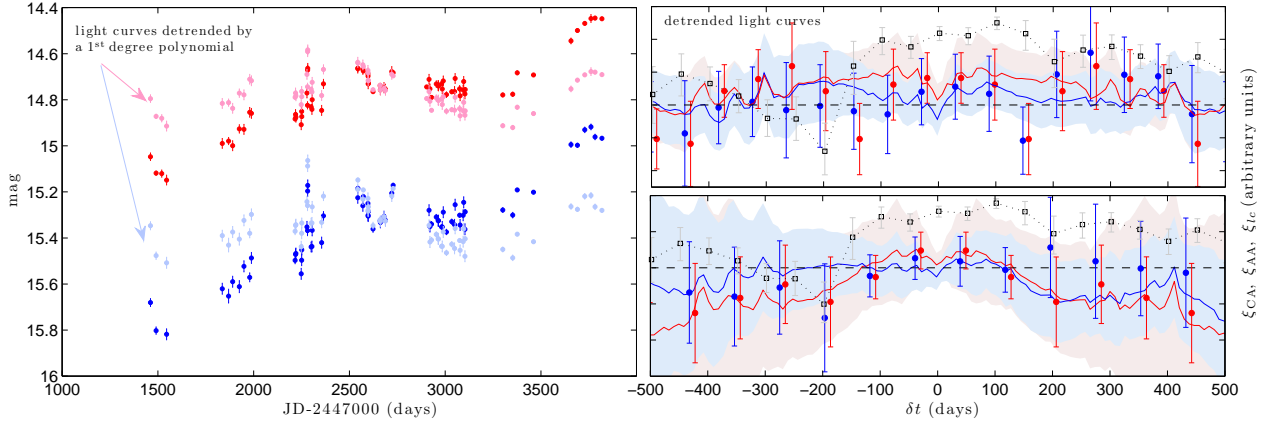}
\caption{Light curve analysis for PG\,0026+129. {\it Left:} The $B$ and $R$ light curves show significant variability on timescales of the order the duration of the light curve. De-trended light curves were calculated by subtracting a $1^{\rm st}$-degree polynomial. The normalized variability measure for the de-trended light curves is a factor two smaller than the original data. {\it Right:} the calculated statistical estimators for the de-trended light curves (upper panel) and the original data (lower panel). See the caption of figure \ref{0804} for the meaning of different curves. While de-trending seems to be able to improve the signal, a statistically significant peak is not detected by either method.}
\label{0026}
\end{figure*}

Like PG\,0804, PG\,0026+129 ($z=0.142$) is among the best-sampled quasars in our sample with $\sim 70$ points per light curve (see Fig. \ref{0026}). While the normalized variability measure is comparable to that of PG\,0804, most of the power is associated with long-term variability. In particular, when subtracting a linear trend from the light curve, the normalized variability measure drops by a factor of $\sim2$ in both bands. At its redshift, the $R$-band is the line-rich (Fig. \ref{wise}) with a net relative line contribution of $\sim 3\%$ due to H$\alpha$ (or $\sim 7$\% if iron is important).

The statistical estimators calculated here do not yield a significant time-lag. Specifically, there is no discernible peak detected in either estimator using either cross-correlation method (i.e., ICCF or ZDCF). While de-trending seems to somewhat improve the signal (there is a $1\sigma$ hump at around 100\,days), it falls short of revealing a 
significant result. We attribute our failure to securely identify a signal to a combination of factors related to the smaller variability measure on the relevant BLR-size timescales, as well as to the lower net contribution of  emission lines to the filters, compared to the case of PG\,0804.

PG\,0052+251 has 72 points in each band, and a normalized variability measure of $26\%/20\%$ in the $B/R$ bands, partly due to a sharp feature extending over $\sim 200$\,days scales. The typical measurement uncertainty is $\lesssim 3\%$ in both bands. At $z=0.155$, the net line contribution is of order 4\%. i.e., comparable to the noise-level. Therefore, despite the relatively large number of visits, no clear signal is detected, in either statistical estimator, using either numerical scheme.

The remaining quasars in the \citet{kas00} PG sample also do not yield a significant signal. This is however expected  given the small number of points in the photometric light curve of these quasars being $\sim$40 (see Fig. \ref{1sigma}).  The fact that such objects do have a spectroscopically-determined time lag is also not surprising: the photometric approach to reverberation mapping is less sensitive than the spectroscopic one. 

\subsubsection{Ensemble averages}

\begin{figure}
\plotone{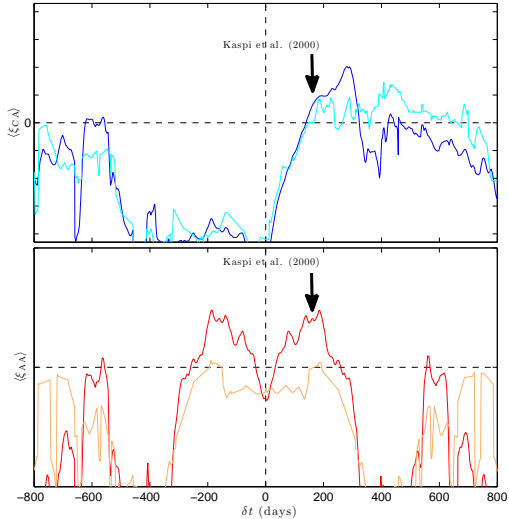}
\caption{Ensemble averages for $\xi_{\rm CA}$ (upper panel) and $\xi_{\rm AA}$ (lower panel), for the PG sample of 17 quasars from \citet[see our table 1]{kas00}. Both the arithmetic mean (dark shades) and the median (light shades) are shown. Ensemble averages for both statistical estimators give consistent results, and peak around 180-280\,days. This is in agreement with the predicted time-lag based on the BLR size vs. luminosity relation of \citet{kas00}, for a typical $L_{45}=1$ quasar (see text). The median shows a less significant peak although the results are qualitatively consistent with the ensemble mean. This demonstrates the feasibility of the photometric reverberation mapping technique in uncovering the typical time-lag in an ensemble of objects for which individual time-lag determination is difficult to achieve.}
\label{pg_mean}
\end{figure}

To overcome the limitations plaguing the analysis of individual objects, ensemble averages may be considered. As demonstrated in \S3, combining the results for many objects, allows the time-lag to be determined with good accuracy, and was a main motivation on our part to develop this approach to reverberation mapping. Clearly, large ensembles are required to reach a robust result when less than favorable light curves are available on an individual case basis. Nevertheless, with upcoming photometric surveys, statistics is unlikely to be a limiting factor, and defining ensemble averages over narrow luminosity and redshift bins will be possible.

As our small sample of 17 PG quasars consists of a range of redshifts and quasar luminosities, we use the following transformation to put the results for different objects on an equal footing: $\delta t \to \delta t/[(1+z)L_{45}^\gamma]$, where $L_{45}\equiv \lambda L_\lambda (5100\,{\rm \AA})/10^{45}\,{\rm erg~s^{-1}}$. This transformation corrects for both cosmological time-dilation, and the fact that the BLR size scales roughly as $L^\gamma$. Following \citet{kas00}, we take $\gamma=0.7$ but note that similar results are obtained for $0.5 \lesssim \gamma \lesssim 0.7$ \citep[and references therein]{ben09}.  Using the above transformation, all PG quasars are effectively transformed to a $z=0$ quasar with $L_{45}=1$. 

Figure \ref{pg_mean} shows the ensemble-averaged results for $\xi_{\rm CA},~\xi_{\rm AA}$ for the entire PG sample (table 1). We deliberately include all objects regardless of data quality issues (as in the case of e.g., PG\,1229). A peak in both (mean) statistical estimators is seen around 200\,days. This value is similar to a time-lag of 170\,days, which is the value predicted by the \citet{kas00} BLR size-luminosity relation for a $L_{45}=1$ quasar. Also plotted are the median results for both statistical estimators. Both the mean and the median provide qualitatively consistent results, although quantitative differences do exist. Given the small nature of the sample, we do not present estimates for the uncertainty on the mean/median values.

To conclude, averaging the results for the PG quasar sample, we obtain a similar peak in both statistical estimators, occurring at times which are qualitatively consistent with the BLR sizes implied by the \citet{kas00} results. A larger sample of objects will allow to better quantify the mean time lag for sub-samples of quasars over narrow luminosity and redshift intervals. Alternatively, better quality data (higher S/N, and better sampled light curves) could be used to better constrain the time-lag for individual objects in the PG sample of quasars.

\section{Summary}

We demonstrated that line-to-continuum time-delays associated with the broad emission line region in quasars can be deduced from broadband photometric light curves. This is made possible by defining appropriate statistical estimators [termed $\xi_{\rm CA}(\delta t)$ and $\xi_{\rm AA}(\delta t)$] to process the data, and which effectively subtract the continuum contribution to the cross-correlation signal of the light curves. Using this formalism, the time-lag is identified with the time $\delta t$ at which these estimators peak. 

The proposed formalism makes use of the fact that line and continuum processes have very different variability properties, and that quasars exhibit large amplitude variations on timescales comparable to the BLR light crossing time. The formalism is generally adequate for cases in which few strong emission lines, having different intensities, contribute to the spectrum, as indeed characterizes the optical emission spectrum of quasars. 

To effectively apply the method, prior knowledge of the quasar redshift (using either photometric or spectroscopic means) is required to identify relatively line-dominated and  line-free spectral bands. Together with knowledge of the filter response functions, these may be used to better implement the algorithm and interpret the results.   Using numerical simulations, we demonstrated that photometric reverberation mapping is feasible under realistic observing conditions and for typical quasar properties. We showed, by means of numerical simulations and real data (pertaining to the PG quasar sample), that the measurement of the line-to-continuum time-delay is robust although somewhat less accurate than that obtained using the spectroscopic reverberation mapping technique. In addition, we quantify biases in the photometrically deduced time-lags, which are sensitive to the BLR geometry and the duration of the light curve. We identify their origin and suggest several methods to correct for them. In addition, we qualitatively study potential biases associated with the presence of multiple emission lines in quasar spectra, and with finite inter-band continuum time-delays. 

The main advantage of the photometric approach to reverberation mapping over other methods is that it is {\it considerably} more efficient and cheap to implement, and can therefore be applied to large samples of objects. In fact, any survey which photometrically monitors the sky with proper cadence and signal-to-noise in two or more filters, and for which quasars may be identified and their redshift estimated, can be used to measure the size of the BLR. We demonstrated that, with large enough samples of quasars, it is possible to beat down the noise, and deduce (statistical) BLR size measurements that are considerably more accurate than are currently achievable by spectroscopic means. 

By combining photometric light curves with single epoch spectroscopic measurements, which allow to estimate the velocity dispersion of the BLR, it may be possible to measure the black hole mass in orders of magnitude more objects than are directly measurable by other means. This would allow for the statistical measurement of the SMBH mass for sub-samples of quasars (selected according to e.g., their luminosity, redshift, or colors), leading to highly accurate results, and potentially revealing differences between different classes of objects, which, thus far, may have eluded detection.

To conclude, photometric surveys that monitor the sky on a regular basis are likely to play a major role in the coming decades in the (statistical) measurement of the BLR size in quasars, as well as in determining their  black hole masses. This will lead to a more complete census of black holes over cosmic time, and shed light on outstanding problems concerning the formation and co-evolution of black holes and their host galaxies.

\acknowledgements

We thank E. Behar, T. Holczer, and S. Kaspi for thought-provoking weekly meetings that triggered much of this work, and the referee for insightful comments. D .C. thanks C. Kochanek, D. Maoz, and H. Netzer for valuable comments on an earlier version of this paper, and S. Kaspi, A. Laor, H. Netzer, and C. Thompson for wise advice.  Fruitful discussions with D. V. Bowen, D. Dultzin, K. Horne, Y. Krongold, S. Rafter, and O. Shemmer are greatly appreciated. We thank M. Vestergaard for providing us with iron UV emission templates. This research has been supported in part by a FP7/IRG PIRG-GA-2009-256434, and by grant 927/11 from the Israeli Science Foundation. E. D. thanks E. Behar for financial support. D. C. thanks Chuli and the people of Arlozorov caf\'e for continuous encouragement.

\end{document}